\documentclass[aps,prd,twocolumn,superscriptaddress,byrevtex,longbibliography,nofootin bib]{revtex4-2}
\usepackage[utf8]{inputenc}
\usepackage{color}
\usepackage{graphicx}
\usepackage{bm}

\usepackage{amsmath,amssymb,amsfonts}

\def\dotmbh{\dot M_{\rm BH}}

\def\mbh{M_{\rm BH}}
\def\mbhtilde{\tilde M_{\rm BH}}
\def\mbhinit{M_{{\rm BH}}(0)}
\def\mbhinittilde{\tilde M_{{\rm BH}}(0)}
\def\mpunc{{\mathcal M}}

\def\mns{M}
\def\mnsinit{M(0)}
\def\mgrav{M}
\def\mrest{M_0}
\def\asym{\star}
\def\dotmbhasym{\dot M_{\rm BH}^\star}
\def\dotmnsasym{\dot M^\star}
\def\dotmbhminasym{\dot M_{\rm BH, min}^\star}
\def\Rinit{R(0)}
\def\flux{{\mathcal F}}
\def\eq{Eq.}
\def\eqs{Eqs.}

\def\rareal{R}
\def\riso{r}
\def\prl{Phys. Rev. Lett.}
\def\prd{Phys. Rev. D}

\def\mnras{Mon. Not. R. Astron. Soc.}

\def\apss{Astrophys. Space Sci.}
\def\jcap{Journal of Cosmology and Astroparticle Physics}
\newcommand\longcomment[1]{}

\begin{document}

\title{Accretion onto a small black hole at the center of a neutron star}

\author{Chloe B.~Richards}

\affiliation{Department of Physics and Astronomy, Bowdoin College, Brunswick, ME 04011}

\author{Thomas W.~Baumgarte}

\affiliation{Department of Physics and Astronomy, Bowdoin College, Brunswick, ME 04011}

\author{Stuart L.~Shapiro}

\affiliation{Department of Physics, University of Illinois at Urbana-Champaign, Urbana, IL 61801}

\affiliation{Department of Astronomy and NCSA, University of Illinois at Urbana-Champaign, Urbana, IL 61801}

\begin{abstract}
We revisit the system consisting of a neutron star that harbors a small, possibly primordial, black hole at its center, focusing on a nonspinning black hole embedded in a nonrotating neutron star.  Extending earlier treatments, we provide an analytical treatment describing the rate of secular accretion of the neutron star matter onto the black hole, adopting the relativistic Bondi accretion formalism for stiff equations of state that we presented elsewhere.  We use these accretion rates to sketch the evolution of the system analytically until the neutron star is completely consumed.  We also perform numerical simulations in full general relativity for black holes with masses up to nine orders of magnitude smaller than the neutron star mass, including a simulation of the entire evolution through collapse for the largest black hole mass.  We construct relativistic initial data for these simulations by generalizing the black hole puncture method to allow for the presence of matter, and evolve these data with a code that is optimally designed to resolve the vastly different length scales present in this problem.  We compare our analytic and numerical results, and provide expressions for the lifetime of neutron stars harboring such endoparasitic black holes.
\end{abstract}

\maketitle

\section{Introduction and Summary}

A number of authors have discussed the prospect of using neutron stars as dark matter detectors (see, e.g., \cite{GolN89,deLF10,BraL14,BraE15,CapPT13,BraLT18,EasL19,GenST20} and references therein).  The idea is that the dark matter may be in the form of small black holes that may be captured by neutron stars, or may be in the form of other particles that may coalesce to form black holes in the interior of neutron stars.  In either case, neutron stars may end up  harboring a small ``endoparasitic" black hole at their center. Such a black hole will then accrete the surrounding neutron star matter until the entire star has been consumed. The current existence of neutron star populations therefore can be used to constrain the nature of black holes as candidates for dark matter.

In principle, the above scenario could be realized in at least two different ways. It is possible that primordial black holes (PBHs) formed in the early Universe (see, e.g., \cite{Haw71,CarH74}), and that they contribute to, or even account for, the dark matter. Such PBHs can be captured by stars.  Following capture, the black holes can settle close to the stellar center, and subsequently accrete the entire star (see, e.g., \cite{Haw71,Mar95}).  The capture, settling and accretion process are particularly efficient for neutron stars (see, e.g., \cite{CapPT13,GenST20}, but see also \cite{MonFVSH19}), so that the current existence of certain neutron star populations constrains the density of PBHs in certain mass ranges. In particular, these arguments have been used to establish limits in the mass range of $10^{-15} M_\odot \lesssim \mbh \lesssim 10^{-9} M_\odot$ \cite{CapPT13}, which is otherwise only poorly constrained (cf.~\cite{KueF17,CarK20,CarKSY20}; see also \cite{SasSTY18,VasV21} and references therein for constraints arising from gravitational wave observations).

Alternatively, it is possible that dark matter particles are captured inside a neutron star.  Under sufficiently favorable conditions, these particles may form a high-density object that may then collapse to form a small black hole (see, e.g., \cite{GolN89,deLF10,BraL14,BraLT18}). The authors of \cite{BraLT18}, for example, estimated the black hole mass to be approximately $\mbh \lesssim 10^{-10} M_\odot$, similar to the mass range of interest for PBHs.

Once a black hole has either been captured by the neutron star, or has formed inside the neutron star, and after this black hole has settled to the center, it will accrete the neutron star matter, ultimately consuming the entire star.  Observational signatures of this process have been discussed by a number of authors, including \cite{FulKT17,TakFK20,GenST20}, while numerical simulations have been performed recently by \cite{EasL19}.  The authors of \cite{EasL19} considered three different equations of state (EOSs), as well as both rotating and non-rotating neutron stars, but limited their simulations to relatively large black hole masses with $\mbh / \mns \geq 10^{-2}$, where $\mns$ is the neutron star mass.  They found that the accretion rate follows the relation $\dotmbh \propto \mbh^2$ as suggested by Bondi accretion (\cite{Bon52}; see also \cite{Shapiro}, hereafter ST, for a textbook treatment, including its relativistic generalization), and that the accretion rate is largely independent of the neutron star spin, in agreement with~\cite{KouT14}. 

In this paper we extend earlier work on the accretion of neutron stars by endoparasitic black holes in several ways. In Section \ref{sec:estimates} 
we provide an analytical overview describing the accretion process in the core of a nonrotating neutron star. Building on our earlier results \cite{RicBS21a,BauS21}, we adopt a relativistic Bondi accretion model to describe a nonspinning black hole accreting gas obeying a stiff, polytropic EOS to track the evolution of the system.  As discussed in \cite{RicBS21a}, Bondi accretion for stiff EOSs with adiabatic indices $\Gamma > 5/3$ exhibits some qualitative differences from corresponding results for soft EOSs; taking these into account we obtain the constant of proportionality in the previously confirmed relation $\dotmbh \propto \mbh^2$.  We integrate the equations, crudely accounting for the quasistatic evolution of the neutron star during the accretion process, to calculate the neutron star's survival time.

In Section \ref{sec:numerics} we perform numerical simulations of the accretion process in full general relativity.  Supported by the results of \cite{EasL19} we focus on nonrotating, spherical neutron stars and nonspinning black holes only, and adopt, for simplicity, a stiff, polytropic EOS with $\Gamma = 2$. Such a value is often used in numerical simulations to approximate a stiff nuclear EOS. Using a numerical relativity code implemented in spherical polar coordinates (see \cite{BauMCM13,BauMM15}) with a logarithmic radial coordinate, we extend the results of \cite{EasL19} by simulating the accretion onto black holes with mass ratios as small as $\mbh / \mns \simeq 10^{-9}$. This allows our simulations to extend into the mass range of interest from the perspective of PBHs and viable black hole dark matter candidates, as discussed above. Such black hole masses thus extend down to the range of dwarf planets.

\begin{figure}[t]
    \centering
    \includegraphics[width = 0.45 \textwidth]{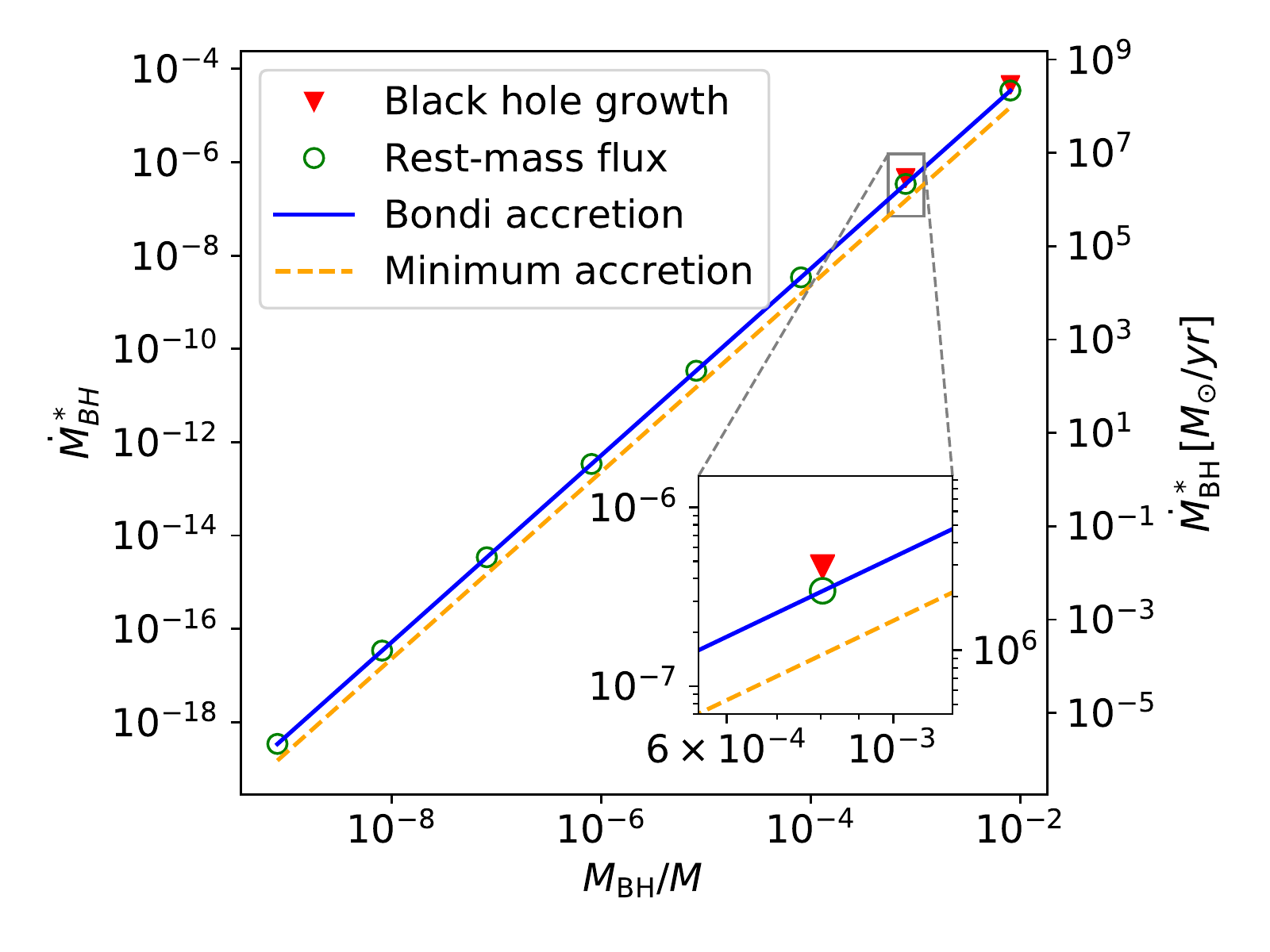} 
    \caption{The analytical and numerical values of the accretion rate $\dotmbhasym$ as a function of $\mbh/\mns$, computed for our fiducial $\Gamma = 2$ neutron star model (see Table \ref{tab:ns} for details).  On the left we express the accretion rate in geometrized units, while on the right we express it in units of solar masses $M_\odot$ per year (since the former is dimensionless, no mass scale is needed for this conversion).  The solid (blue) line represents the analytical solution (\ref{mdot_fidu}) based on Bondi accretion for stiff EOSs, while (red) triangles and (green) circles represent numerical measurements of the accretion rate based on the growth of the black hole horizon and on the rest-mass flux, respectively (see Sections \ref{sec:numerics:diagnostics:bh_growth} and \ref{sec:numerics:diagnostics:flux}).  The dashed (red) line represents the minimum accretion rate (\ref{mdot_min}) for a $\Gamma = 2$ polytrope.  (See also Table \ref{tab:accretion} for a detailed listing of these results.)} 
    \label{fig:accretion_rates}
\end{figure}

We compare our analytical and numerical results in Section \ref{sec:results}.  In particular, we compile accretion rates obtained both analytically and numerically for a large range of black hole masses in Table \ref{tab:accretion}.  We also plot these results in Fig.~\ref{fig:accretion_rates}, which serves as a summary of our results.  In this figure, the solid line represents the analytical accretion rate quoted in Eq.~(\ref{mdot_fidu}). The dashed line shows the {\it minimum} accretion rate for $\Gamma = 2$ as identified by \cite{RicBS21a} (see their \eq~51). Such a minimum only arises for stiff EOSs, but is surprisingly insensitive to the EOS for $\Gamma >2$.  The open circles and filled triangles mark two different methods for computing the accretion rate in our numerical simulation data.  The former result from evaluating the accretion rate directly from the flux of matter crossing the black hole horizon, while the latter are based on measuring the rate at which the area of the horizon increases. Since the black hole grows at a rate that is proportional to the square of the black hole mass, and hence is exceedingly slow for small black hole masses, we can accurately measure the horizon's growth rate numerically only for the largest black hole masses.  We note that the proportionality $\dotmbh \propto \mbh^2$ holds over many orders of magnitude, that our numerically determined accretion rates agree well with those computed analytically from the relativistic Bondi accretion relation (indicating that we have correctly identified the constant of proportionality), and finally that the actual accretion rates are only slightly larger than the minimum accretion rate identified in \cite{RicBS21a}.  All of the above confirms that the consumption of a neutron star by an endoparasitic black hole is well described by Bondi accretion for stiff EOSs, so that these analytical expressions can be used to establish the lifetime of such neutron stars (see Section \ref{sec:estimates:times}).

Throughout this paper we use geometrized units with $G = 1 = c$ unless noted otherwise.

\section{Analytical Overview}
\label{sec:estimates}

\subsection{Accretion rates}
\label{sec:estimates:rates}

We assume that a nonspinning black hole with mass $\mbh$ is located at the center of a spherical 
neutron star with mass $\mns \gg \mbh$ and radius $R \gg \mbh$ that would be in strict hydrostatic equilibrium in the absence of the black hole.  We take the matter to be at rest far from the black hole, which resides at the center of a nearly homogeneous core. To estimate the rate at which the black hole accretes matter from the neutron star we first define the capture radius
\begin{equation} \label{ra}
r_a \equiv \frac{\mbh}{a_c^2}.    
\end{equation}
Here $a$ is the speed of sound, given by
\begin{equation} \label{a}
a^2 = \left. \frac{dP}{d \rho} \right|_s = \left. \frac{dP}{d \rho_0} \right|_s \, \frac{\rho_0}{\rho + P},
\end{equation}
and $a_c$ is its value evaluated at the center of the unperturbed star.   In (\ref{a}), $P$ is the pressure, $\rho$ the total mass-energy density, $\rho_0$ the rest-mass density, and the derivative is taken at constant entropy.  For the Newtonian estimates that we will provide in this section we will not need to distinguish between $\rho$ and $\rho_0$, but we introduce them here for later reference.  For a polytropic equation of state with
\begin{equation} \label{polytrope}
P = K \rho_0^{\Gamma},
\end{equation}
where $\Gamma$ is the adiabatic index and $K$ a constant,
we may then approximate
\begin{equation} \label{ac}
a_c \simeq \left( \frac{\Gamma P_c}{\rho_c} \right)^{1/2} 
\simeq \left( \frac{\Gamma \mns}{R} \right)^{1/2}.
\end{equation}  
Here $P_c$ and $\rho_c$ are the values of the unperturbed star's pressure and density at the center,
with $P_c \ll \rho_c$ and where we have assumed hydrostatic equilibrium in writing the last (rough) equality.

The accretion flow onto the black hole can be approximated in two opposite limiting regimes, depending on whether the neutron star mass $m(r_a)$ contained within the capture radius $r_a$ is greater or smaller than the black hole mass $\mbh$.  In the latter case we may ignore the self-gravity of the neutron star matter, so that the accretion process is described by secular Bondi accretion, i.e.~adiabatic flow with asymptotically constant matter density and pressure and zero flow velocity (\cite{Bon52,Mic72}; see also ST for a textbook treatment).  In the former case we cannot ignore the self-gravity, and the evolution modeled as an accretion process 
becomes catastrophic dynamical collapse.

Defining $m(r)$ as the neutron star mass within the radius $r$, so that $\mns = m(R)$, we have
\begin{equation} \label{m(r)}
m(r) \simeq \frac{4 \pi}{3} \rho_c r^3
\end{equation}
for sufficiently small $r$. We then compute the 
crucial mass ratio, 
\begin{equation} \label{mass_ratio1}
\frac{m(r_a)}{\mbh} \simeq \frac{4 \pi}{3} \, \frac{\mbh^2 \rho_c}{a_c^6} \simeq \frac{4 \pi}{3} \, \frac{\mbh^2 \rho_c R^3}{\Gamma^3 \mns^3}.
\end{equation}
We now write
\begin{equation} \label{rhoc}
\rho_c = \delta \, \bar \rho = \delta \, \frac{3 \mns}{4 \pi R^3},
\end{equation}
where $\bar \rho$ is the unperturbed star's mean density, and the factor $\delta$ measures its central concentration, $\rho_c / \bar \rho$.  We tabulate values of $\delta$ for Newtonian polytropes of index $n$ and adiabatic index $\Gamma = 1 + 1/n$ in Table \ref{tab:polytropes}.  Inserting (\ref{rhoc}) into (\ref{mass_ratio1}) we now have
\begin{equation} \label{mass_ratio}
\frac{m(r_a)}{\mbh} \simeq \frac{\delta}{\Gamma^3} \, \left( \frac{\mbh}{\mns} \right)^2.
\end{equation}
We see that $m(r_a) \sim \mbh$ only for large black hole masses $\mbh \sim \mns$ and soft EOSs that result in large values of $\delta/\Gamma^3$, also listed in Table \ref{tab:polytropes}. For black holes that start out with masses much smaller than the neutron star mass, $\mbh \ll \mns$, almost the entire accretion process (i.e.~the longest duration) will occur in the regime $m(r_a) \ll \mbh$, and will therefore be described by quasistatic Bondi accretion, while only the short final epoch proceeds dynamically in the regime $m(r_a) \sim \mbh$.  In the following two Sections we will provide estimates for these cases separately, namely for Bondi accretion in Section \ref{sec:estimates:bondi} and for dynamical accretion in Section \ref{sec:estimates:dynamical}.

\begin{table}[t]
    \centering
    \begin{tabular}{c|c|c|c|c}
         $n$ &  $\Gamma$ & $\delta$ & $\delta/\Gamma^3$ & $\lambda_s$ \\ 
         \hline
         3.0~ & ~4/3~ & ~54.2~ & ~22.94~ & ~0.707 \\ 
         2.5~ & ~7/5~ & ~23.3~ & ~8.45~ & ~0.625 \\
         2.0~& ~3/2~ & ~11.4~ & ~3.38~ & ~0.500 \\
         1.5~ & ~5/3~ & ~5.99~ & ~1.296~ & ~0.250 \\ 
         1.0~ & ~2.0~ & ~3.29~ & ~0.411~ & -- \\
         0.5 & 3.0 & 1.84 & 0.068 & -- 
    \end{tabular}
    \caption{Values of the central condensation $\delta$ as well as the combination $\Gamma^3 / \delta$ for Newtonian polytropes with polytropic index $n$ and adiabatic index $\Gamma = 1 + 1/n$.  For $\Gamma \leq 5/3$ we also include the accretion eigenvalues $\lambda_s$; see Eq.~(\ref{lambda_s}). For $\Gamma > 5/3$, the relativistic accretion eigenvalues $\lambda_{\rm GR}$ can be constructed as in \cite{RicBS21a}; see also Section \ref{sec:estimates:fid} below.}
    \label{tab:polytropes}
\end{table}

\subsubsection{Case I: $m(r_a) \ll \mbh$ -- Bondi accretion}
\label{sec:estimates:bondi}

In this case we can neglect the self-gravity of the neutron star fluid inside the capture radius $r_a$, since the gravitational forces are dominated by those exerted by the black hole.  In this case, the accretion is well described by adiabatic Bondi accretion \cite{Bon52}, the rate for which is given by 
\begin{equation} \label{mdot_bondi}
\dotmbhasym = - \dotmnsasym = 4 \pi \lambda_{\rm GR} \left( \frac{\mbh}{a_\asym^2} \right)^2 \rho_\asym a_\asym.
\end{equation}
Here $\lambda_{\rm GR}$ is a dimensionless ``accretion eigenvalue", typically of order unity, and the $\asym$ symbols denote values as observed by a ``local asymptotic" static observer who is far from the black hole, but still well inside the neutron star, i.e.~$\mbh \ll r \ll R$. The dot in the accretion rates denotes a derivative with respect to time as measured by such an observer (see also Appendix \ref{app:schwarzschild} for a detailed discussion). It is assumed that the density $\rho_\asym$ and sound speed $a_\asym$ approach constants, and the flow speed $u_\asym$ approaches zero, as $r > r_a$ becomes large in this asymptotic region, which typically resides inside the nearly homogeneous core of the neutron star. In Appendix \ref{app:schwarzschild}, as well as Section \ref{sec:numerics:diagnostics:bh_growth} below, we discuss how this ``local" accretion rate is related to the rate of mass accretion as seen by an observer far from the star. Here we simply point out that the accretion of bound matter is not influenced by the spherical matter distribution beyond the radius at which it becomes bound (Birkhoff's theorem): when self-gravity of the matter bound to the black hole can be neglected, the only gravitational source influencing the accretion is the central black hole.

The accreted mass measured by (\ref{mdot_bondi}) is fundamentally a (baryon) {\it rest} mass.  It enhances the black hole's total gravitational mass by a similar amount for strictly adiabatic flow as long as the asymptotic internal energy of the gas is small in comparison to the rest-mass energy (see Section \ref{sec:results:rates} and Table \ref{tab:accretion}).  We will ignore this difference in the context of the approximate treatment in this Section. 

A relativistic treatment of accretion for nonspinning black holes shows that the requirement that the sound speed be less than the speed of light demands that the flow become transonic and pass through a critical point, yielding a unique value for $\lambda_{\rm GR}$, as shown by ST. For soft equations of state with $\Gamma \leq 5/3$ and $a_\asym \ll 1$ these accretion eigenvalues can be found from a Newtonian treatment of Bondi accretion, resulting in
\begin{equation} \label{lambda_s}
\lambda_{\rm GR} \rightarrow \lambda_s = \frac{1}{4} \left( \frac{5 - 3 \Gamma}{2} \right)^{(3 \Gamma - 5)/2(\Gamma - 1)}  
\end{equation}
(see, e.g., Chapter 14, Eq.~(14.3.17), and Table 14.1 in ST).  Unlike $\lambda_{\rm GR}$, the values of $\lambda_s$ are independent of $a_\asym$ and $a_s$, the sound speed at the transonic radius $r_s$ (see \cite{RicBS21a} for details).  Stiffer equations of state require a relativistic treatment as discussed in \cite{RicBS21a}. 

We now assume that $r_a$ is sufficiently small, $r_a \ll R$, that we can approximate the fluid variables as seen by the local observer discussed above, i.e.~$\rho_\asym$ and $a_\asym$, by those at the center of the unperturbed star, i.e.~$\rho_c$ and $a_c$. The accretion rate then becomes
\begin{equation}
\label{acc0}
\dotmbhasym = - \dotmnsasym = 4 \pi  \lambda_{\rm GR} \frac{\mbh^2}{a_c^3} \rho_c,
\end{equation}
from which we can crudely estimate the accretion 
timescale $\tau_{\rm acc}$ 
\begin{equation} \label{tau_acc_bondi0}
\tau_{\rm acc} \equiv \frac{\mbh}{\dotmbhasym} \simeq \frac{a_c^3}{ 4 \pi \lambda_{\rm GR} \mbh \rho_c} 
= \frac{\Gamma^{3/2}}{3\delta \lambda_{\rm GR}} \, \frac{\mns^{1/2} R^{3/2}}{\mbh}.
\end{equation}
Here we have used (\ref{ac}) and (\ref{rhoc}) in the last step.  The above estimates the timescale for the black hole to double its mass, which is the bottleneck in the neutron star consumption process: this epoch takes the longest because both the black hole mass and hence the accretion rate~(\ref{acc0}) are the smallest they will be during the process. Dividing by the neutron star mass, we may rewrite this result as
\begin{equation} \label{tau_acc_bondi}
\frac{\tau_{\rm acc}}{\mns} \simeq \frac{\Gamma^{3/2}}{3\delta \lambda_{\rm GR}} \, \left( \frac{R}{\mns} \right)^{3/2} \, \left( \frac{\mns}{\mbh} \right).
\end{equation}
Note that $\tau_{\rm acc}/\mns \rightarrow \infty$ as $\mbh / \mns \rightarrow 0$.  Alternatively, we may also express the accretion timescale in terms of the neutron star's dynamical (collapse) timescale
\begin{equation} \label{tau_dyn}
\tau_{\rm dyn} \simeq \frac{\gamma}{(4 \pi \rho_c / 3)^{1/2}} 
= \frac{\gamma}{\delta^{1/2}} \left( \frac{R}{\mns} \right)^{3/2} \mns,
\end{equation}
where $\gamma$ is a factor of order unity and where we have used (\ref{rhoc}).  Combining (\ref{tau_acc_bondi0}) and (\ref{tau_dyn}) we obtain
\begin{equation} \label{tau_acc_bondi2}
    \frac{\tau_{\rm acc}}{\tau_{\rm dyn}} \simeq
    \frac{\Gamma^{3/2}}{\delta^{1/2} \lambda_{\rm GR} \gamma} \,
    \left( \frac{\mns}{\mbh} \right).
\end{equation}
We again have $\tau_{\rm acc}/\tau_{\rm dyn} \rightarrow \infty$ as $\mbh / \mns \rightarrow 0$.  We will calculate $\tau_{\rm acc}$ more carefully in Section \ref{sec:estimates:times} below.

\subsubsection{Case II: $m(r_a) \sim \mbh$ -- Dynamical accretion}
\label{sec:estimates:dynamical}

In this case we cannot neglect the self-gravity of the neutron star fluid inside the capture radius $r_a$.  We now generalize the definition (\ref{ra}) of this capture radius, and define a critical radius,
\begin{equation} \label{rcrit}
    r_{\rm crit} = \frac{m(r_{\rm crit}) + \mbh}{a_c^2},
\end{equation}
inside which the gas is marginally bound by the combined mass of the black hole and the gas itself.  Using (\ref{m(r)}), we can rewrite (\ref{rcrit}) as
\begin{equation} \label{temp1}
    \frac{4 \pi}{3} \rho_c r_{\rm crit}^2 
    \left( 1 + \frac{\mbh}{m(r_{\rm crit})} \right) = a_c^2.
\end{equation}
The rate at which the black hole accretes mass can now be expressed as the area of the sphere with the critical radius, $4 \pi r_{\rm crit}^2$, times the mass flux across this sphere, $\rho_c u_c$.  Assuming that, at the critical radius, the fluid speed $u_c$ is comparable to the sound speed $a_c$, as in typical Bondi flows, we then have
\begin{equation}
    \dotmbhasym \simeq 4 \pi r_{\rm crit}^2 \rho_c a_c 
    = 3 a_c^3 \left( 1 + \frac{\mbh}{m(r_{\rm crit})} \right)^{-1},
\end{equation}
where we have used (\ref{temp1}) in the last equality.  The corresponding accretion timescale is then given by
\begin{equation}
    \tau_{\rm acc} = \frac{\mbh}{\dotmbh} \simeq \frac{\mbh}{3 \Gamma^{3/2}} \, \left( \frac{R}{M} \right)^{3/2}  \left( 1 + \frac{\mbh}{m(r_{\rm crit})} \right),
\end{equation}
or
\begin{equation} \label{tau_acc_dyn}
    \frac{\tau_{\rm acc}}{\tau_{\rm dyn}} \simeq \frac{\delta^{1/2}}{3 \gamma \Gamma^{3/2}} \, \frac{\mbh}{\mns} \,
    \left( 1 + \frac{\mbh}{m(r_{\rm crit})} \right),
\end{equation}
where we have approximated the dynamical timescale $\tau_{\rm dyn}$ as in (\ref{tau_dyn}).

We now evaluate \eq~(\ref{tau_acc_dyn}) in two limits.  In the limit $\mbh \sim m(r_{\rm crit})$, in which case $\mbh \sim \mns$ by (\ref{mass_ratio}), we notice that the accretion timescale $\tau_{\rm acc}$ becomes comparable to the dynamical timescale $\tau_{\rm dyn}$, as one would expect. In the opposite limit, $\mbh \gg m(r_{\rm crit})$, the critical radius $r_{\rm crit}$ defined in (\ref{rcrit}) reduces to $r_a$ defined in (\ref{ra}) and we may approximate
\begin{equation} \label{mass_ratio2}
    \frac{\mbh}{m(r_{\rm crit})} \simeq \frac{3 \mbh}{4 \pi \rho_c r_a^3} = \frac{3 a_c^6}{4 \pi \rho_c \mbh^2} = 
    \frac{\Gamma^3}{\delta} \, \left( \frac{\mns}{\mbh} \right)^2, 
\end{equation}
where we have used (\ref{ac}) and (\ref{rhoc}) in the last equality.  Inserting (\ref{mass_ratio2}) into (\ref{tau_acc_dyn}) we recover, up to factors of order unity, the Bondi accretion timescale (\ref{tau_acc_bondi2}), as expected.

\subsection{Effects of stellar evolution}
\label{sec:estimates:evolution}

Our simple estimates (\ref{tau_acc_bondi2}) and (\ref{tau_acc_dyn}) for the accretion timescales ignore the fact that the accretion rates change as the black hole mass $\mbh$ increases, and also ignore the fact that the neutron star structure changes as the accretion proceeds.  We can approximate the effects of this secular ``stellar evolution" by assuming that, while the star loses mass to the black hole, it will adjust quasistatically to a new equilibrium configuration while keeping its total Newtonian energy $E$ constant. We now write this energy as the simple functional
\begin{equation} \label{energy}
     E = -\alpha\frac{\mns \mbh}{R} - \frac{3\Gamma - 4}{5\Gamma - 6} \,\frac{\mns^2}{R},
\end{equation}
where the first term accounts for the interaction between the stellar gas and the black hole, with $\alpha$ being a constant that depends on $\Gamma$, $\alpha = \alpha(\Gamma)$, while the second term describes the neutron star's self-energy
(see \eq~3.3.10 in ST).

Evaluating (\ref{energy}) at the initial time, denoted by $(0)$, we have 
\begin{equation}  \label{energy_initial}
     E = -\alpha\frac{\mnsinit \mbhinit}{\Rinit} - \frac{3\Gamma - 4}{5\Gamma - 6} \,\frac{\mnsinit^2}{\Rinit}.
\end{equation}
Since, by our assumption, expressions (\ref{energy}) and (\ref{energy_initial}) must be identical, we can equate them and solve for $R$ to find
\begin{equation} \label{R_NS1}
R = \frac{\mns}{\mnsinit} \,
\frac{\alpha \mbh + (3 \Gamma - 4) \mns /(5 \Gamma - 6)}{\alpha \mbhinit + (3 \Gamma - 4) \mnsinit /(5 \Gamma - 6)} \,
\Rinit. 
\end{equation}
We now approximate $\mbh \ll \mns$, in which case (\ref{R_NS1}) reduces to
\begin{equation} \label{R_NS}
R \simeq \left( \frac{\mns}{\mnsinit} \right)^{2}\, \Rinit.
\end{equation}
Using (\ref{R_NS}) in (\ref{ac}) then yields
\begin{equation} \label{ac2}
    a_c \simeq \left( \frac{\Gamma \mnsinit}{\Rinit} \right)^{1/2} \, \left( \frac{\mnsinit}{\mns} \right)^{1/2},
\end{equation}
while (\ref{rhoc}) gives
\begin{equation} \label{rhoc2}
\rho_c \simeq \delta \, \frac{3 \mnsinit}{4 \pi \Rinit^2}
\left( \frac{\mnsinit}{\mns} \right)^5.
\end{equation}
Inserting (\ref{ac2}) and (\ref{rhoc2}) into the Bondi accretion rate (\ref{mdot_bondi}) then results in
\begin{equation} \label{mdot_bondi_evolve}
    \dot M = - \frac{3 \lambda_{\rm GR} \, \delta}{\Gamma^{3/2}}
    \, \frac{(\mbhinit + \mnsinit - \mns)^2}{\mnsinit^{1/2} \Rinit^{3/2}} \, \left( \frac{\mns}{\mnsinit} \right)^{-7/2} 
\end{equation}
where we have expressed the black-hole mass $\mbh$ in terms of the evolving neutron star mass $\mns$ as
\begin{equation}\label{mass_conserv}
    \mbh = \mbhinit + \mnsinit - \mns.
\end{equation}  
Note that the last factor in (\ref{mdot_bondi_evolve}) accounts for stellar evolution.

\subsection{Accretion times}
\label{sec:estimates:times}

We can now compute the neutron star lifetime (i.e.~the accretion time) by integrating 
\eq~(\ref{mdot_bondi_evolve}).  Towards that end, it is useful to introduce the dimensionless quantities
\begin{equation} \label{yfac}
y_{0} \equiv \frac{\mbhinit + \mnsinit}{\mnsinit},~~~~~
y \equiv \frac{\mns}{\mnsinit},
\end{equation}
and
\begin{equation}
    \label{Tfac}
    T \equiv \frac{3 \lambda_{\rm GR}\, \delta}{\Gamma^{3/2}} \left( \frac{\mnsinit}{\Rinit^3} \right)^{1/2} t, 
\end{equation}
in terms of which we may rewrite (\ref{mdot_bondi_evolve}) as
\begin{equation} \label{dydT}
    \frac{dy}{d T} = - (y_0 - y)^2 \, y^{-7/2}.
\end{equation}
As in (\ref{mdot_bondi_evolve}), the last factor accounts for stellar evolution.

\subsubsection{Without stellar evolution}
\label{sec:estimates:times:noevolve}

We first ignore the effects of stellar evolution, so that (\ref{dydT}) reduces to
\begin{equation}
     \frac{dy}{d T} = - (y_0 - y)^2,
\end{equation}
which can be integrated readily to yield
\begin{equation}
\left[ T \right]_i^f = \left[\frac{1}{y - y_0} \right]_i^f.
\end{equation}
Here the square brackets serve as a reminder to insert limits of integration.  At the initial time, which we choose to be $T_i = 0$, we have $y_i = 1$, so that
\begin{equation} \label{T_no_evolve_1}
    T_f = \frac{1}{y_f - y_0} - \frac{1}{1 - y_0}
    = \frac{\mnsinit}{\mbhinit} - \frac{\mnsinit}{\mbh}
\end{equation}
where $\mbh$ is the black hole mass at the time $T_f$, and where we have used (\ref{mass_conserv}).  We can now find the total accretion time by setting the final neutron star mass equal to zero, i.e.~by choosing $y_f = 0$.  Further assuming that $\mbhinit \ll \mnsinit$
and recalling (\ref{Tfac}) we find
\begin{equation} \label{t_no_evolve}
\frac{\tau_{\rm acc}}{\mnsinit} = \frac{\Gamma^{3/2}}{3 \delta \lambda_{\rm GR}} \left( \frac{\Rinit}{\mnsinit} \right)^{3/2} \left( \frac{\mnsinit}{\mbhinit} \right),
\end{equation}
which is identical to (\ref{tau_acc_bondi}), as expected.

\subsubsection{With stellar evolution}
\label{sec:estimates:times:evolve}

We now repeat the exercise, but include the last factor in (\ref{mdot_bondi_evolve}) in order to account for stellar evolution.  In this case the integral can be carried out as described in Appendix \ref{sec:integral}.  Choosing, as before, $y_i = 1$ at $T_i = 0$, as well as $y_f = 0$ in order to obtain the accretion time $T_f = T_{\rm acc}$, we obtain 
\begin{equation} \label{T_evolve_1}
T_{\rm acc} =  
\frac{5}{2} + \frac{4}{3} y_0 + 6 y_0^2 + \frac{y_0^3}{y_0 - 1} - 7 y_0^{5/2} \ln \left( \frac{y_0^{1/2} + 1}{y_0^{1/2} - 1} \right).
\end{equation}
 Alternatively, we may introduce
\begin{equation}
    y_{h0} = y_0 - 1 = \frac{\mbhinit}{\mnsinit}
\end{equation}
and rewrite (\ref{T_evolve_1}) as
\begin{eqnarray} \label{T_evolve}
T_{\rm acc}
& = &  7y_{h0}^2 + \frac{49}{3}y_{h0} + \frac{161}{15}  + \frac{1}{y_{h0}} \nonumber \\ 
& & -\frac{7}{2}(y_{h0} + 1)^{5/2}\ln{\frac{\sqrt{1+y_{h0}}+1}{\sqrt{1+y_{h0}}-1}}.
\end{eqnarray}
 
Taking the limit $y_{h0} \rightarrow 0$, we see that $T_{\rm acc}$ will be dominated by the term $1 / y_{h0}$, so that we recover the same accretion time $t_{\rm acc}$ as in (\ref{t_no_evolve}).\footnote{Note, however, that we have modeled the effects of stellar evolution to leading order only, see Section \ref{sec:estimates:evolution}, so that only the leading-order corrections to our results in the absence of stellar evolution have physical significance.}  This is not entirely surprising, since most of the accretion time is spent during early times when neither the neutron star mass nor radius change appreciably, so that stellar evolution is not important. At late times, however, the response of the star to the accretion process, and the corresponding adjustments in the stellar structure, will affect the accretion time, as expressed by (\ref{T_evolve}).

As a concrete example, consider a star with $M = 1 M_{\odot}$.
Since, in geometrized units, $1 M_\odot \simeq 1.4 \mbox{~km} \simeq 5~\mu\mbox{s}$, we then have
\begin{equation}
\tau_{\rm acc} \sim \left( \frac{\Rinit}{\mnsinit} \right)^{3/2} \left( \frac{\mnsinit}{\mbhinit} \right) 10^{-5} \mbox{~s},
\end{equation}
where we ignored factors of order unity.  For a main-sequence star, with $R \simeq  10^5 M$, we see that the accretion time will exceed a Hubble time if $\mbhinit \lesssim 10^{-15} M_\odot$.  For a neutron star, however, $R \simeq 10 M$, resulting in significantly smaller accretion timescales.  Therefore, black holes with masses as small as $\mbhinit \gtrsim 10^{-21} M_\odot$ would be able to consume a neutron star well within a Hubble time.

\subsection{Fiducial neutron star model}
\label{sec:estimates:fid}

\begin{table*}[]
    \centering
    \begin{tabular}{c|c|c|c|c}
        Quantity & 
        Rescaled wrt\footnote{With respect to}~$K$ &
        Rescaled wrt~$\mgrav$ &
        Rescaled wrt~max.~mass model &
        Physical units
        \\
        \hline
         $\rho_{0c}$\footnote{Central rest-mass density} & 
         $\tilde \rho_{0c} = 0.2$ &
         $\mgrav^2 \rho_{0c} = 0.00495$ &
         $\rho_{0c}/\rho_{0c}^{\rm max} = 0.629$ &          
         $\rho_{0c} = 3.41 \times 10^{15}$ g/cm$^3$ 
         \\
         $\rho_c$\footnote{Central mass-energy density} &
         $\tilde \rho_{c} = 0.24 $ &
         $\mgrav^2 \rho_{0c} = 0.0059$ &
         $\rho_{c}/\rho_{c}^{\rm max} = 0.572$ &
         $\rho_{c} = 4.09 \times 10^{15}$ g/cm$^3$
         \\
         $R$~\footnote{Areal radius} &
         $\tilde R = 0.865$ & 
         $R/\mgrav = 5.50$ & 
         $R/R^{max} = 1.13$ &
         $R = 10.8$ km 
         \\
         $r_{\rm iso}$\footnote{Isotropic radius} &
         $\tilde r_{\rm iso} = 0.699$ &
         $r_{\rm iso}/M = 4.45$ & 
         $r_{\rm iso}/r_{\rm iso}^{\rm max} = 1.19$ & 
         $r_{\rm iso} = 8.73$ km 
         \\
         $\mgrav$~\footnote{Gravitational mass} & 
         $\tilde \mgrav = 0.157$ &
         $\mgrav/\mgrav = 1$ & 
         $\mgrav/\mgrav^{\rm max} = 0.959$ &
         $M = 2.80 \times 10^{33}$ g 
         \\
         $\mrest$\footnote{Rest mass} &
         $\tilde \mrest = 0.176$ & 
         $\mrest/\mgrav = 1.12$ & 
         $\mrest/\mrest^{\rm max} = 0.954$ &
         $\mrest = 3.14 \times 10^{33}$ g 
         \\
         $\psi_c$\footnote{Central conformal factor} &
         $\psi_c = 1.27$ & 
        $\psi_c = 1.27$ &  
        $\psi_c/\psi_c^{\rm max} = 0.933$  &
        $\psi_c = 1.27$ \\
        $\alpha_c$\footnote{Central lapse function} &
        $\alpha_c = 0.570$ &  
        $\alpha_c = 0.570$ & 
        $\alpha_c/\alpha_c^{\rm max} = 1.23 $ &
        $\alpha_c = 0.570$ 
    \end{tabular}
    \caption{ Parameters for our fiducial $\Gamma = 2$, $n=1$ polytropic neutron star model in the absence of a black hole (see text for details).  The conformal factor $\psi$ is well defined by our assumption that, in Cartesian coordinates, the determinant of the conformally related metric is $\bar \gamma = 1$.  The lapse function $\alpha$ listed here is the value obtained from integrating the TOV equations, and is different from the 1+log lapse adopted in our numerical evolution calculations (see \eq~\ref{1+log}). }
    \label{tab:ns}
\end{table*}

In our comparisons with the numerical results of Section \ref{sec:numerics} below we consider, as a fiducial neutron star model, a dynamically stable equilibrium star with a central rest-mass density of $\rho_{0c} = 0.2\, K^{-n}$ governed by a polytropic equation of state (\ref{polytrope}) with $\Gamma = 2$ and $n = 1$, which we constructed by solving the Tolman-Oppenheimer-Volkoff (TOV) \cite{Tol39,OppV39} equations.   Detailed properties of this stellar model are listed in Table \ref{tab:ns}.

Since $K^{n/2}$ has units of length in geometrized units, we may introduce non-dimensional quantities by rescaling any dimensional quantity with a suitable power of $K$; in particular, we define
\begin{equation}
    \begin{array}{rclrcl}
        \tilde \rho & \equiv & K^n \rho, & 
        \tilde R & \equiv K^{-n/2} R, \\
        \tilde \rho_0 & \equiv & K^n \rho_0, ~~~~&
        \tilde \mgrav & \equiv K^{-n/2} \mgrav,
    \end{array}
\end{equation}
and similar for other quantities.  We list ``tilde" variables that have been rescaled with respect to $K$ in the second column in Table \ref{tab:ns}.  In the third column we rescale each variable with respect to the neutron star's gravitational mass $\mgrav$, while in the fourth column we rescale with respect to the corresponding maximum-mass configuration.  In particular we note that, for our adopted model, $\mgrav/\mgrav^{\rm max} = 0.959$, where $\mgrav^{\rm max}$ is the maximum gravitational mass of a spherical star with our adopted EOS.  Finally, for the fifth column in Table \ref{tab:ns} we assume that our star has a gravitational mass of $\mgrav = 1.4 \, M_\odot$, in which case $K$ takes the value $K = (1.4 \, M_\odot / \tilde \mgrav)^2 \simeq 156$~km$^2$. 

From the parameters given in Table \ref{tab:ns} we compute the central sound speed to be
\begin{equation}
    a_c = 0.534.
\end{equation}
We identify, as before, the neutron star's central density and sound speed with the corresponding asymptotic values for the Bondi accretion onto the black hole and follow \cite{RicBS21a} to compute the accretion eigenvalue
\begin{equation}
    \lambda_{\rm GR} = 1.29.
\end{equation}
Inserting the above values into the Bondi accretion rate (\ref{mdot_bondi}) we obtain
\begin{equation} \label{mdot_fidu}
    \dotmbhasym  = 21.24 \,\mbhtilde^2, 
\end{equation}
which is not significantly larger than the {\it minimum} steady state accretion rate for a stiff polytrope with $\Gamma = 2$, 
\begin{equation} \label{mdot_min}
    \dotmbhminasym = 9.29 \, \mbhtilde^2
\end{equation}
(see \eq~51 in \cite{RicBS21a}).  Adopting the above value of $K$, and recalling that, in geometrized units, $M_\odot \simeq 5 \times 10^{-6}$~s, we can evaluate (\ref{mdot_min}) to yield   
\begin{equation}
\dotmbhminasym = 7.33 \times 10^{-9}\, \frac{M_\odot}{\mbox{yr}} \left( \frac{\mbh}{10^{-10} M_\odot} \right)^2.
\end{equation}

\section{Numerical treatment}
\label{sec:numerics}

\subsection{Initial data}
\label{sec:numerics:indata}

We construct relativistic initial data describing a nonspinning black hole embedded at the center of a nonrotating, spherical neutron star. Our task is to solve the Hamiltonian and momentum constraints of general relativity (see, e.g. \cite{BauS10} for discussion and references) which we do by generalizing the puncture method (see \cite{BraB97}) to allow for the presence of matter.  Our approach differs from that adopted by \cite{EasL19}, who constructed initial data by matching an interior black hole solution to an exterior neutron star solution.

We assume that the initial slice is conformally flat, so that we may write the spatial metric as $\gamma_{ij} = \psi^4 \, \eta_{ij}$, where $\psi$ is the conformal factor and $\eta_{ij}$ the flat metric.  We also assume the initial slice to be momentarily static, and thereby choose the extrinsic curvature to vanish, $K_{ij} = 0$, and the initial momentum density measured by a normal observer to vanish, $S_i \equiv - \gamma_{ia} n_b T^{ab} = 0$.  Here a ``normal observer" is an observer whose four-velocity is the normal vector $n^a$ on the spatial slice, and $T^{ab}$ is the stress-energy tensor.  With these assumptions the momentum constraints are satisfied trivially, and the Hamiltonian constraint becomes
\begin{equation} \label{Ham_1}
\bar D^2 \psi = - 2 \pi \psi^5 \rho,
\end{equation}
where $\bar D^2$ is the flat Laplace operator, and $\rho \equiv n_a n_b T^{ab}$ the mass-energy density as observed by a normal observer. We allow for a conformal rescaling of the density,
\begin{equation}\label{conformal_recale}
\rho = \psi^m \bar \rho,
\end{equation}
where $m$ is a yet-to-be-determined exponent.  An attractive choice might be $m = -6$, since it leaves the proper integral over the density $\rho$ invariant if we keep $\bar \rho$ fixed,
\begin{equation}
\int \psi^6 \rho \,d^3 x = \int \bar \rho \, d^3 x.
\end{equation}
Note, however, that this integral represents neither the gravitational nor the rest mass.

Now assume that we have constructed a solution to the TOV equations \cite{Tol39,OppV39} in isotropic coordinates, so that we obtain radial profiles of the conformal factor $\psi_{\rm NS}$ and the mass-energy density $\rho_{\rm NS}$ for the equilibrium neutron star by itself.   In particular, these functions satisfy the Hamiltonian constraint (\ref{Ham_1}) with
\begin{equation} \label{Ham_NS}
\bar D^2 \psi_{\rm NS} = - 2 \pi \psi_{\rm NS}^5 \rho_{\rm NS},
\end{equation}
and we identify
\begin{equation}
\bar \rho = \psi_{\rm NS}^{-m} \, \rho_{\rm NS}.
\end{equation}
We now want to modify this solution so that the new solution accounts for a black hole embedded at the center of the neutron star.  Towards that end, we write the conformal factor as a sum of contributions from the neutron star and the black hole,  as well as a correction $u$,
\begin{equation} \label{psi_ansatz}
\psi = \psi_{\rm NS} + \psi_{\rm BH} + u
\end{equation}
and recast the Hamiltonian constraint as an equation for $u$.  In (\ref{psi_ansatz}) we have introduced
\begin{equation}
\psi_{\rm BH} = \frac{\mpunc}{2 r}
\end{equation}
as an isolated black hole's contribution to the conformal factor in our isotropic coordinates.  We refer to $\mpunc$ as the ``puncture mass"; it has no immediate physical significance, and serves as a mass parameter only (see Section \ref{sec:numerics:diagnostics:mass} and \eqs~\ref{m_irr} and \ref{m_irr_approx} below for the black hole's isolated horizon, or irreducible, mass).  Inserting (\ref{psi_ansatz})  into the Hamiltonian constraint (\ref{Ham_1}) and observing that $\bar D^2 \psi_{\rm BH} = 0$ we obtain
\begin{equation}
\bar D^2 \psi_{\rm NS} + \bar D^2 u = - 2 \pi \left( \psi_{\rm NS} + \psi_{\rm BH} + u \right)^{5 + m} \bar \rho,
\end{equation}
or, using (\ref{Ham_NS}),
\begin{equation} \label{Ham_u}
\bar D^2 u = - 2 \pi \left\{ \left( \psi_{\rm NS} + \psi_{\rm BH} + u \right)^{5 + m} - \psi_{\rm NS}^{5 + m} \right\} \bar \rho.
\end{equation}
Since $\psi_{\rm BH}$ diverges as $r \rightarrow 0$, it may be desirable to choose $m < -5$, which makes the right-hand side of (\ref{Ham_u}) regular.  Therefore, unless noted otherwise, we will use $m = -6$ in all our simulations.  In Appendix \ref{sec:app:approx} we derive an approximate but analytical 
solution to (\ref{Ham_u}), and discuss some of the properties of the solutions $u$ (see also Fig.~\ref{fig:u_approx}).

We have now reduced the problem to finding regular solutions $u$ to the elliptic equation (\ref{Ham_u}), subject to the Robin boundary condition $u \propto 1/r$ for large $r$.  Since the equation is non-linear, we adopt an iterative approach.  Once we have obtained this solution we can compute the new (physical) energy density $\rho$ from
\begin{equation}
\rho = \psi^m \bar \rho = \left( \frac{\psi_{\rm NS} + \psi_{\rm BH} + u}{\psi_{\rm NS}} \right)^m \rho_{\rm NS}.
\end{equation}
Note that we will have $\rho \rightarrow 0$ as $r \rightarrow 0$ with $m < 0$ initially.

\begin{figure}
    \centering
    \includegraphics[width = 0.45 \textwidth]{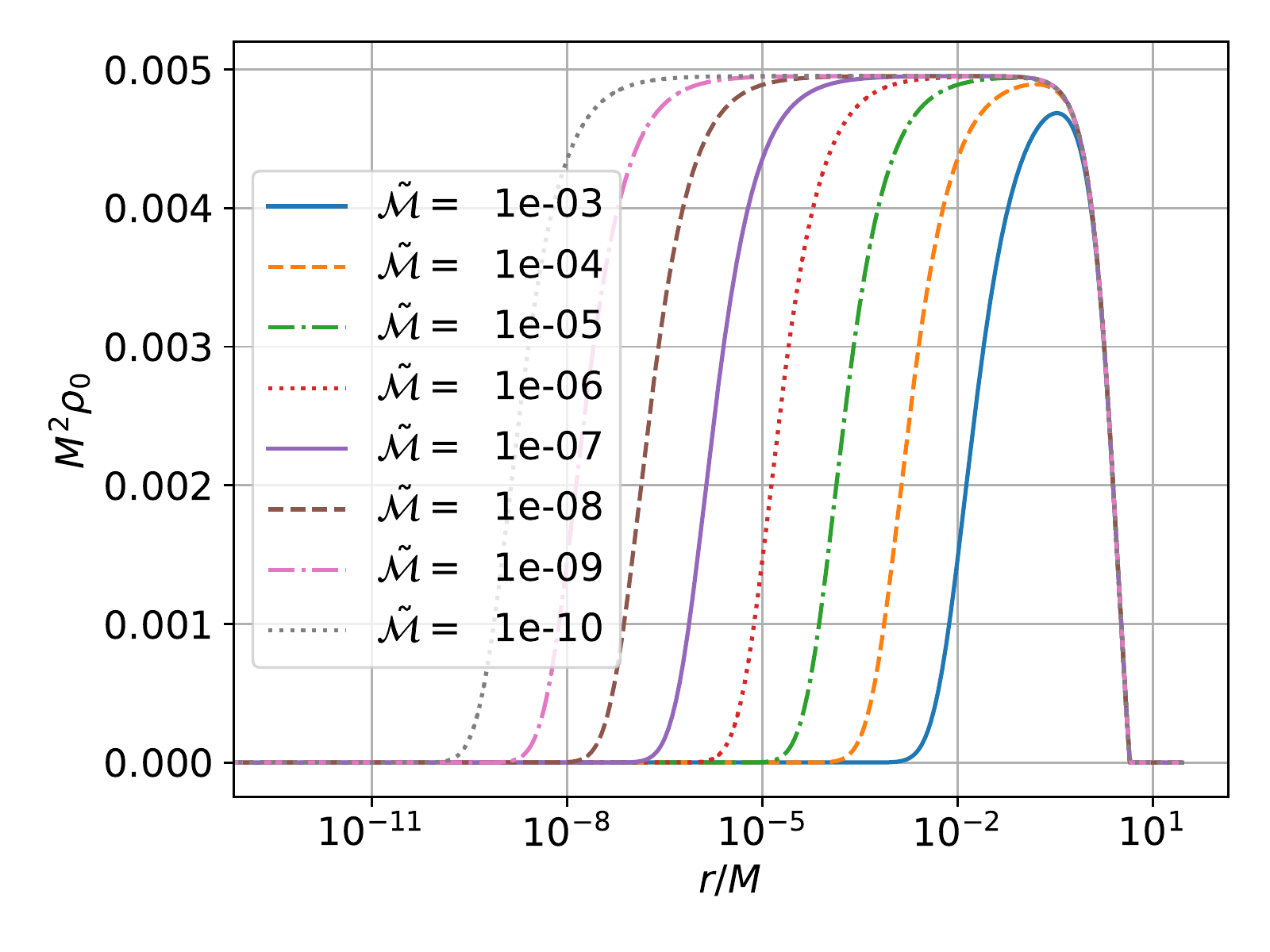} 
    \caption{Profiles of the initial rest-mass density $\rho_0$ as a function of isotropic radius $r$, for our fiducial neutron star model (see Table \ref{tab:ns}) with different black hole puncture masses $\mpunc$.  Here and throughout we choose $m = -6$ in (\ref{conformal_recale}) unless stated otherwise.  Even for tiny black hole masses, the logarithmic radial variable allows us to resolve the vastly different length scales of the black hole and the neutron star.}
    \label{fig:rho_0_diff_mass}
\end{figure}

We have implemented the above approach in the code described in \cite{BauMCM13,BauMM15}, which solves Einstein's equations in spherical polar coordinates.  We use a logarithmic radial coordinate, which allows us to resolve both the black hole and the neutron star with modest numerical resources, even when $\mpunc \ll \mns$ (see Section \ref{sec:numerics:evolution} for details).  We show examples of density profiles for black holes with different masses embedded in our fiducial neutron star model in Fig.~\ref{fig:rho_0_diff_mass}.

While our initial data depend on our choice of $m$ in \eq~(\ref{conformal_recale}), they quickly settle into a quasiequilibrium configuration soon after matter marginally bound to the black hole begins to flow inward at 
$t \sim r_a/a_c \sim 7 \mbh$. 
The system thus relaxes to a state of quasistatic accretion onto the black hole that is independent of $m$ (see Figs.~\ref{fig:rho_0_rays} 
as well as the discussion in \ref{sec:numerics:evolution} below).

\subsection{Numerical evolution}
\label{sec:numerics:evolution}

We evolve our initial data with a code that solves the Baumgarte-Shapiro-Shibata-Nakamura (BSSN) formulation of Einstein's equations \cite{NakOK87,ShiN95,BauS99}.  We adopt a reference-metric formulation (see, e.g.,~\cite{BonGGN04,ShiUF04,Bro09,Gou12}) in order to implement the equations in spherical polar coordinates (see \cite{BauMCM13,BauMM15} for details and tests; see also \cite{MilB17} for tests with Bondi accretion as well as \cite{RucEB18,MewZCREB18,MewZCBEAC20} for other implementations of this approach).  The latest version of our code uses fourth-order finite differencing for all spatial derivatives in Einstein's equations, together with a fourth-order Runge-Kutta time integrator.  

We impose coordinates using the ``1+log" slicing condition
\begin{equation} \label{1+log}
    (\partial_t - \beta^i \partial_i)\, \alpha = - 2 \alpha K 
\end{equation}
(see \cite{BonMSS95}) for the lapse function $\alpha$, and a ``Gamma-driver" condition for the shift vector $\beta^i$ (see \cite{Alcetal03,ThiBB11}).  On our initial slice we choose a ``pre-collapsed" lapse with $\alpha~=~\psi^{-2}$ and zero shift.

For all simulations reported in this paper we used a numerical grid of $N_r = 512$ radial grid points with the outer boundary at $\tilde r_{\rm out} = 4$, corresponding to about 5.7 times the isotropic radius of our fiducial neutron star.  We allocate the radial grid points using a sinh function, resulting in a grid that becomes logarithmic asymptotically and allows us to resolve the vastly different length scales associated with the black hole and the neutron star.  We adjust the parameters of this sinh function for each black hole mass so that the black hole is resolved by approximately 50 grid points, and the smallest grid spacing (at the center of the black hole) is approximately 1\% of the black hole's isotropic radius or less.

We similarly implement the equations of relativistic hydrodynamics in spherical polar coordinates adopting a reference-metric formulation \cite{MonBM14}.  We solve the resulting equations using a Harten-Lax-van Leer-Einfeldt (HLLE) approximate Riemann solver \cite{HarLL83,Ein88}, together with a simple monotonized central-difference limiter reconstruction scheme \cite{Van77}.
We solve these equations adopting an ideal gas law
\begin{equation} \label{gammalaw}
    P = (\Gamma - 1) \rho_0 \epsilon,
\end{equation}
where $\epsilon$ is the specific internal energy density, in terms of which the total mass energy density is given by $\rho = \rho_0 (1 + \epsilon)$.  While \eq~(\ref{gammalaw}) allows for non-isentropic flow, e.g.~shocks, we have found in our numerical simulations that the relation between $P$ and $\rho_0$ remains very close to the polytropic relation (\ref{polytrope}), indicating that the accretion flow is laminar and nearly adiabatic.  As before we focus on $\Gamma = 2$ and we refer to \cite{EasL19} for a survey of different EOSs.  We also note that, for stiff EOSs with polytropic index $0.5 \leq n \leq 1.5$ (or $5/3 \leq \Gamma \leq 3$), there exists a maximum accretion timescale that is nearly independent of the polytropic index;  see \cite{BauS21}.

\begin{figure}[t]
    \centering
    \includegraphics[width = 0.45 \textwidth]{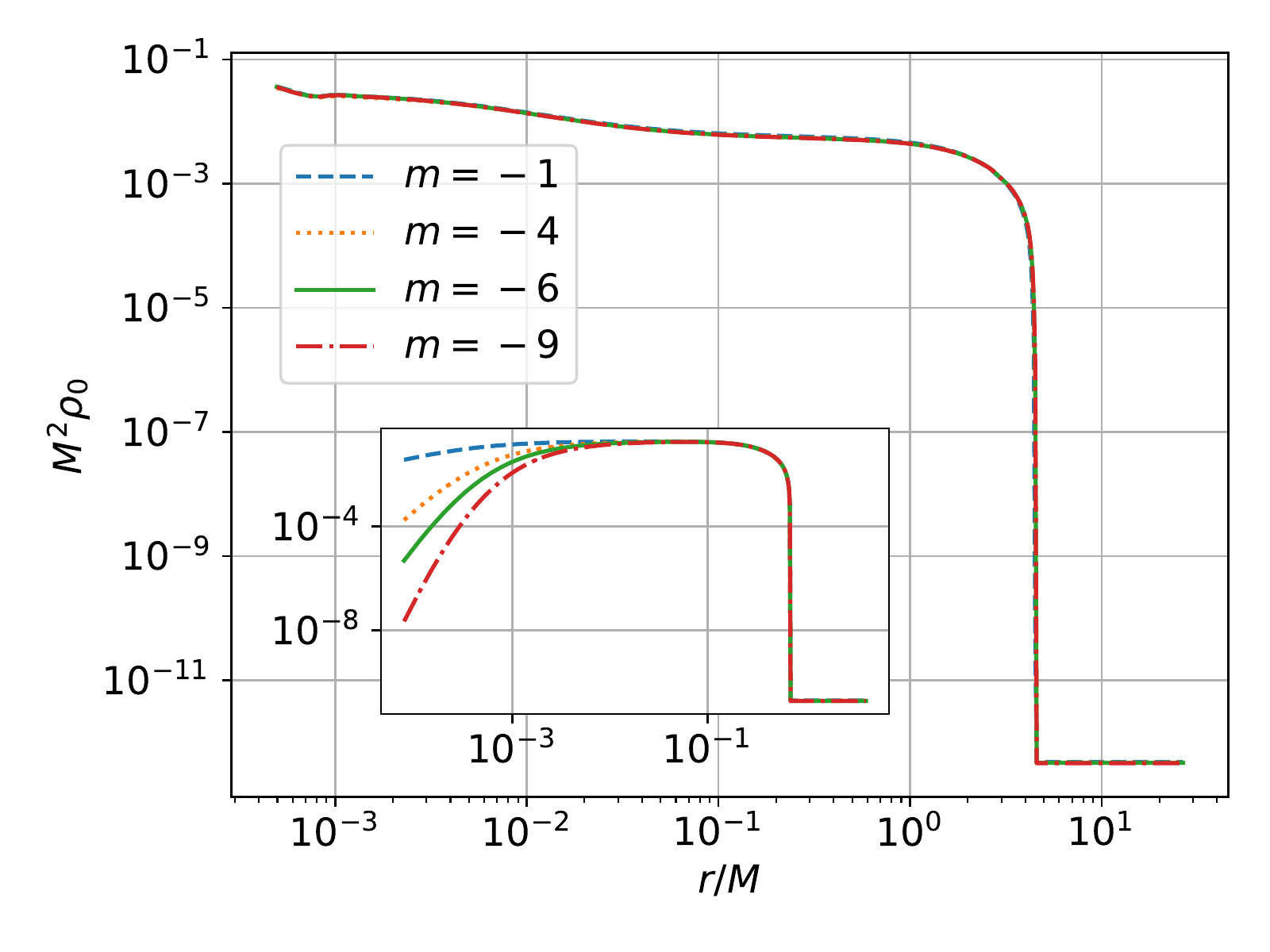} 
    \caption{The rest-mass density $\rho_0$ as a function of isotropic radius $r$ for different values of the conformal exponent $m$ in (\ref{conformal_recale}) for $\tilde \mpunc = 10^{-3}$, so $\mbhinittilde = 1.267 \times 10^{-3}$, and  $\mbhinit / \mnsinit = 8.03 \times 10^{-3}$.  Although the initial  density profiles, shown in the inset, clearly depend on $m$, they all evolve to the same density profile, shown in the large plot, once a quasi-equilibrium has been reached. Here and in several of the following graphs we suppress the innermost few grid points well inside the horizon, since they are affected by numerical noise caused by the puncture singularity at the origin.}
    \label{fig:rho_0_rays}
\end{figure}

As an example of our evolution calculations, we show in Fig.~\ref{fig:rho_0_rays} profiles of the density $\rho_0$ for our fiducial neutron star model hosting a black hole with puncture mass $\tilde \mpunc = 1 \times 10^{-3}$.  We show results for different values of the conformal exponent $m$ in (\ref{conformal_recale}).  Evidently, the initial density profiles, shown in the inset of Fig.~\ref{fig:rho_0_rays}, show large differences, as one might expect.  However, once the evolution reaches quasi-equilibrium, shown in the main graph in Fig.~\ref{fig:rho_0_rays}, the profiles are all very similar.  This gives us confidence that, except for a small initial transient, our evolution calculations are largely independent of our choice of $m$, and quickly settle down into a solution describing a steady-state accretion onto the endoparasitic black hole. 

\subsection{Diagnostics}
\label{sec:numerics:diagnostics}

We invoke a number of different diagnostics in order to evaluate our numerical simulations.

\subsubsection{Black hole Mass}
\label{sec:numerics:diagnostics:mass}

A black hole's isolated horizon, or irreducible, mass is given by
\begin{equation} \label{m_irr}
    \mbh = \left( \frac{\mathcal A}{16 \pi} \right)^{1/2},
\end{equation}
where $\mathcal A$ is the proper area of the black hole's event horizon at a given instant of time.  

In practice, we locate in our numerical evolution calculations the black hole's apparent horizon rather than the event horizon, since the former requires data at one instant of time only, rather than the entire spacetime (see, e.g., \cite{BauS10} for a textbook discussion).   The two horizons should coincide for quasistatic evolution.  We then compute the apparent horizon's proper area, and use this value in (\ref{m_irr}).  For many situations, in particular for stationary or nearly stationary spacetimes, this yields an excellent approximation.  

We can also compute the approximate initial black-hole mass as follows.  Since our initial data are conformally flat and describe a moment of time-symmetry with zero shift, we may write the spacetime metric at that instant as
\begin{equation}
    ds^2 = - \alpha^2 dt^2 + \psi^4 (dr^2 + r^2 d\Omega^2).
\end{equation}
The expansion of a bundle of outgoing null geodesics orthogonal to a spherical surface of radius $r$ is then given by
\begin{equation} \label{expansion}
    \Theta = \frac{\sqrt{2}}{r \psi^2} \frac{d}{dr} \left( r \psi^2 \right)
\end{equation}
(see, e.g., \eq~(7.22) in \cite{BauS10} with $A^2 = B^2 = \psi^4$).  We will assume that $u$ in (\ref{psi_ansatz}) is small compared to the other terms (see Appendix \ref{sec:app:approx}) so that we may approximate
\begin{equation} \label{psi_drop_u}
    \psi \simeq \psi_{\rm NS} + \psi_{\rm BH} = \psi_{\rm NS} + \frac{\mpunc}{2 r}.
\end{equation}
For $r$ much smaller than the neutron star radius we may approximate $\psi_{\rm NS}$ as constant.  

We now find the black hole's apparent horizon by setting the expansion (\ref{expansion}) to zero, which yields
\begin{equation}
    \frac{d}{dr} \left( r \left(\psi_{\rm NS} + \frac{\mpunc}{2r} \right)^2 \right) \simeq \psi_{\rm NS}^2 - \frac{\mpunc^2}{4 r^2} = 0
\end{equation}
or
\begin{equation} \label{r_AH}
    r_{\rm AH} \simeq \frac{\mpunc}{2 \psi_{\rm NS}}.
\end{equation}
At the apparent horizon, we then have
\begin{equation} \label{psi_AH}
    \psi_{\rm AH} \equiv \psi(r_{\rm AH}) \simeq 2 \psi_{\rm NS}.
\end{equation}
We now compute the apparent horizon's proper area from
\begin{equation} \label{area_AH}
    {\mathcal A} = 4 \pi \psi^4_{\rm AH} r_{\rm AH}^2
    = 16 \pi \psi_{\rm NS}^2 \mpunc^2,
\end{equation}
where have used both (\ref{r_AH}) and (\ref{psi_AH}).  Inserting (\ref{area_AH}) into (\ref{m_irr}) we obtain our result
\begin{equation} \label{m_irr_approx}
    \mbh \simeq \psi_{\rm NS} \mpunc,
\end{equation}
where $\psi_{\rm NS}$ may be estimated by the central value of the unperturbed neutron star's conformal factor. We have found that, for $\mpunc \ll \mns$, \eq~(\ref{m_irr_approx}) provides an excellent approximation to our numerical values for the initial black hole masses (see Table \ref{tab:accretion}). 


\subsubsection{Accretion rates: black hole growth}
\label{sec:numerics:diagnostics:bh_growth}

We measure the rates at which the black hole accretes neutron star matter in two different ways (see also Fig.~\ref{fig:accretion_rates} and Table \ref{tab:accretion}).

\begin{figure}[t]
    \centering
    \includegraphics[width = 0.45 \textwidth]{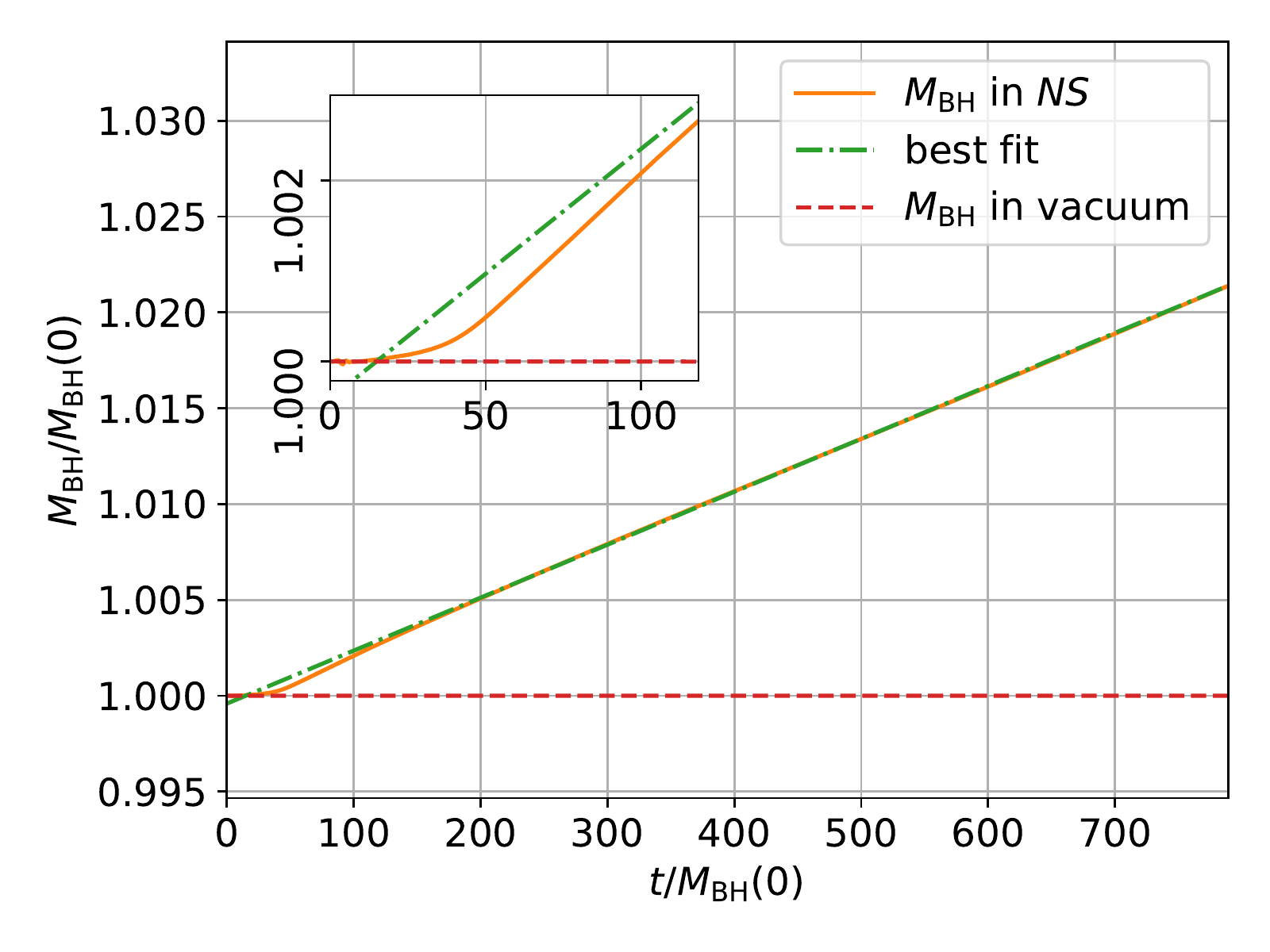} 
    \caption{Growth of the black hole's irreducible mass (\ref{m_irr}) as a function of time, for $\mbhinittilde = 1.26 \times 10^{-3}$ (the solid orange line).  The dashed-dotted (green) line shows a linear fit, whose slope we identify with the accretion rate $\dotmbh$.  For comparison, we also include as the dashed (red) line the irreducible mass of a black hole with the same initial mass $\mbhinit$ in vacuum, i.e.~without a neutron star.}
    \label{fig:BH_vs_NS_BH}
\end{figure}

In one approach, we directly measure the black hole mass $\mbh$ from (\ref{m_irr}) as a function of coordinate time, and then determine the accretion rate from the slope of this function.  We show an example in Fig.~\ref{fig:BH_vs_NS_BH}.  We also included in Fig.~\ref{fig:BH_vs_NS_BH} results for a black hole of the same initial mass $\mbhinit$, but evolved in vacuum, i.e.~without a neutron star, shown as the dashed (red) line.  The latter appears nearly horizontal, demonstrating that the growth observed for a black hole embedded in a neutron star indeed results from accretion of neutron star material, rather than numerical noise.  For small black hole masses, however, the accretion becomes so slow (cf.~\eq~\ref{mdot_bondi}) that it is no longer possible to accurately determine the slope of the function $\mbh(t)$.  Therefore we have used this direct measure of the black hole growth only for black holes with $\mbhinittilde \gtrsim 10^{-4}$ (see Fig.~\ref{fig:accretion_rates} and Table \ref{tab:accretion}).

Measuring the slope of curves $\mbh(t)$ yields the accretion rate $\dotmbh$, where the derivative is taken with respect to the coordinate time $t$.  Since this coordinate time agrees with proper time of a static observer at infinity, i.e.~one at large distances from the neutron star with $ r \gg R$, this measure determines the accretion rate as observed by a static observer at infinity.  In Section \ref{sec:estimates:bondi}, we introduced the accretion rate $\dotmbhasym$ as observed by a static ``local asymptotic" observer far from the black hole, but well inside the star, with $\mbh \ll r \ll R$.  In $\dotmbhasym$, the derivative is therefore taken with respect to this local observer's proper time $\tau_\asym$ (see also Appendix \ref{app:schwarzschild} for a detailed discussion). We can relate the two rates by recognizing that the proper time of the static, local observer moving along a normal vector to our spacelike hypersurfaces  will advance at a rate $d\tau_\asym = \alpha_\asym dt$, where $\alpha_\asym$ is the lapse function of this local observer.  We then have
\begin{equation} \label{relate_rates}
    \dotmbh = \frac{d \mbh}{dt} 
    = \alpha_\asym \frac{ d \mbh}{d \tau_\asym} =
    \alpha_\asym \dotmbhasym
\end{equation}
(see also \eq~\ref{relate_rates_app}).  We may interpret this relation as stating that the rate as observed by a distant observer, $\dotmbh$, is red-shifted by the lapse function $\alpha_\asym$ with respect to the rate as observed by a local observer, $\dotmbhasym$, as one might expect. 

\begin{figure}[t]
    \centering
    \includegraphics[width = 0.45 \textwidth]{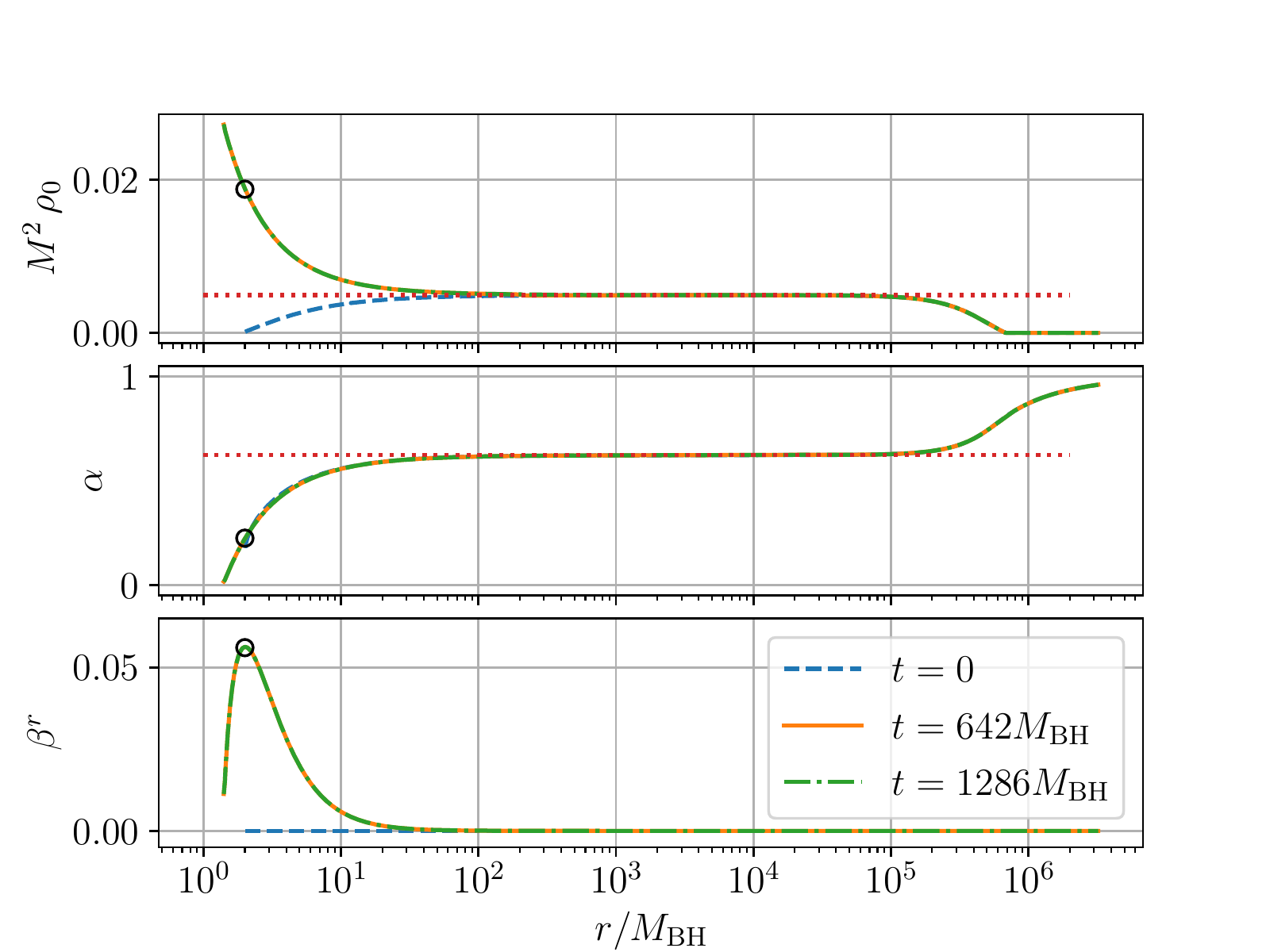} 
    \caption{Profiles of the rest-mass density $\rho_0$, lapse $\alpha$ and shift vector $\beta^r$ in an evolution of our fiducial neutron star with a black hole of mass $\mbhinittilde = 1.267 \times 10^{-6}$ at different (coordinate) times.  At the later times, all functions have settled down into a (quasi) equilibrium.  Note also that the density and lapse ``plateau" in a region $\mbh \ll r \ll R$, where they take the nearly constant values $\rho_{0\asym} \simeq \rho_{0c} = 0.2$ and $\alpha_\asym \simeq 0.623$  (marked by the horizontal dotted lines).  Note also that the shift is very close to zero in this region.  The black circles denote the location of the black hole horizon.}
    \label{fig:star_profile}
\end{figure}

In Fig.~\ref{fig:star_profile} we show profiles of the rest-mass density $\rho_0$, the lapse function $\alpha$ and the shift vector $\beta^r$ in one of our simulations.  This example shows how the density and lapse function ``plateau" in a region $\mbh \ll r \ll R$, so that their values for a ``local asymptotic" observer can be identified to high accuracy as long as $\mbh \ll \mns$.  Note also that, in this region, the shift $\beta^r$ is very close to zero, so that in this nearly static spacetime a static observer (i.e.~one whose four-velocity is aligned with the timelike Killing vector) indeed coincides with a normal observer, as we had assumed above.

We list results for accretion rates determined numerically from the growth of the black hole mass in Table~\ref{tab:accretion}.

\subsubsection{Accretion rates: rest-mass flux}
\label{sec:numerics:diagnostics:flux}

\begin{figure}[t]
    \centering
    \includegraphics[width = 0.45 \textwidth]{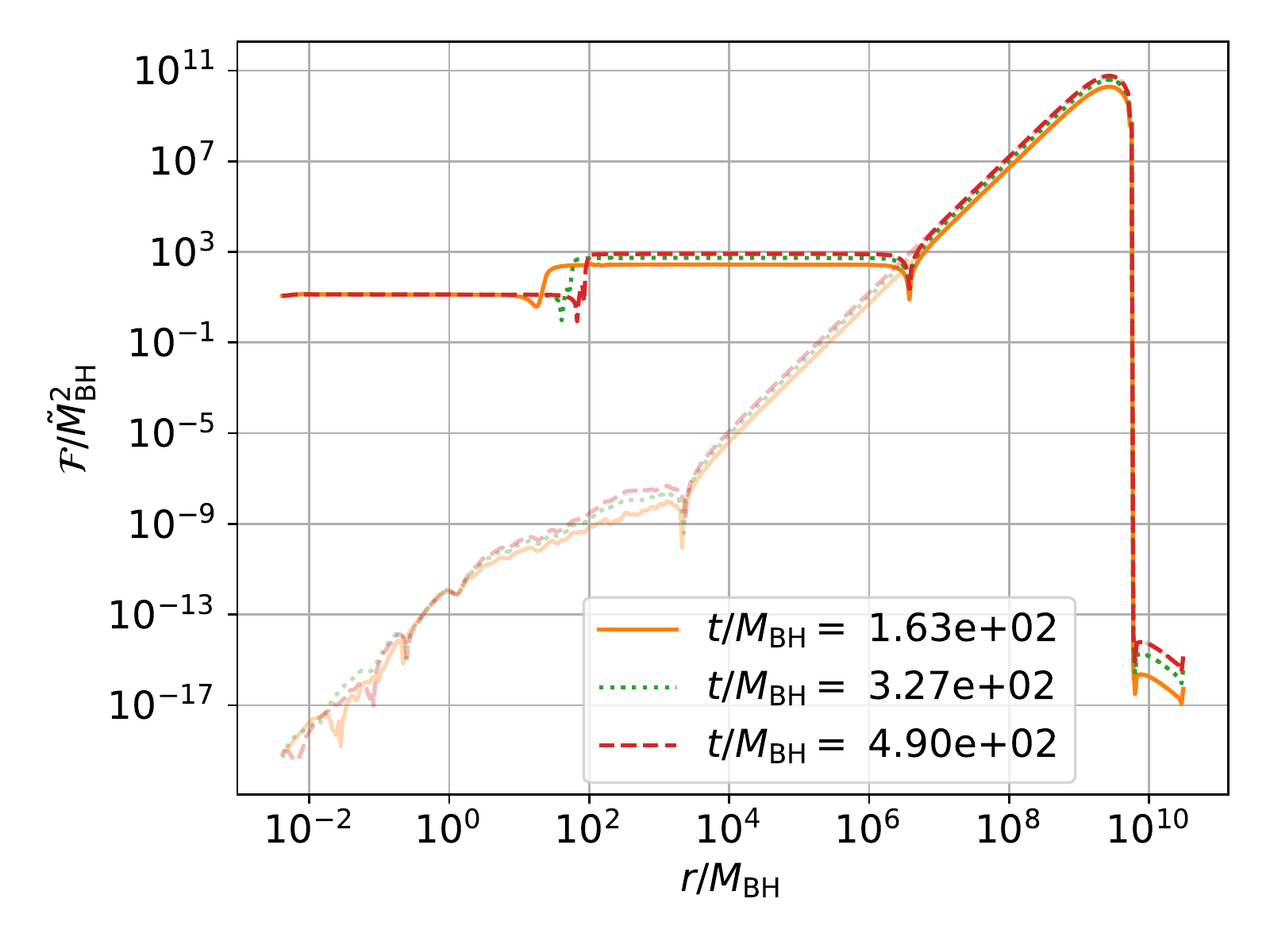} 
    \caption{Profiles of the flux (\ref{flux}) at different instants of time, for a black hole with initial mass $\mbhinittilde = 1.267 \times 10^{-10}$, $\mpunc = 1 \times 10^{-10}$ embedded in our fiducial neutron star model.  In the outer parts of the star, the non-zero flux reflects a numerical adjustment of the star, resulting from the fact that the initial data are not in perfect equilibrium on the numerical grid.  In the inner part, the flux approaches a value that becomes independent of both space and time, resulting in an equilibrium accretion flow onto the black hole.  For comparison we also included, as the faint lines, profiles for an evolution of the same neutron star but without a black hole, which shows the same behavior in the outer parts of the star, but very different behavior in the vicinity of the black hole.}
    \label{fig:acc_ray_m_neg10}
\end{figure}

An alternative approach to computing the accretion rate is to measure the rate of fluid flow across the black hole horizon.  Assuming that the accretion is sufficiently slow so that we can approximate the black hole as nearly static, we can compute the flux $\flux$ of rest-mass accretion through a sphere ${\mathcal H}$ of radius $r$ from
\begin{equation} 
\flux(r) = - \int_{\mathcal H} \sqrt{-g} \rho_0 u^r d\theta d \phi,
\end{equation}
where $g$ is the determinant of the spacetime metric (see, e.g., Appendix A in \cite{FarLS10}).  Assuming spherical symmetry we may carry out the integration to obtain
\begin{equation} \label{flux}
\flux(r) = - 4 \pi \alpha \sqrt{\gamma} \rho_0 u^r r^2,
\end{equation}
where we have used $\sqrt{-g} = \alpha \sqrt{\gamma}$, and where $\gamma$ is the determinant of the spatial metric.  This expression yields the flux of rest-mass through any sphere of radius $r$, and the accretion rate in particular when evaluated on the black hole horizon,
\begin{equation} \label{mdot_flux}
    \dotmbh = \flux(r_{\rm hor}).
\end{equation}
For stationary flow we expect $\flux$ to become independent of radius.  We demonstrate this behavior in Fig.~\ref{fig:acc_ray_m_neg10}, where we show profiles of $\flux$ at different instants of time, for a black hole with initial mass $\mbhinit/\mns = 7.12 \times 10^{-10}$  embedded in our fiducial neutron star.  Note that $\flux$ takes a nearly constant value in an inner region that grows with time, as the fluid flow settles down into steady-state accretion onto the black hole.

Also included in Fig.~\ref{fig:acc_ray_m_neg10} are profiles of the flux $\flux$ for an evolution of our fiducial neutron star model without a black hole.  While these profiles show the same behavior as the neutron star with the black hole in the outer parts of the star, where the flux is dominated by a numerical adjustment of the near-equilibrium initial data to the numerical grid, the profiles are very different in the vicinity of the black hole.  This demonstrates that the plateau in the flux observed for the evolution with the black hole indeed represents steady-state accretion onto the black hole, rather than a numerical artifact.

We record our numerical results for these accretion rates in Table \ref{tab:accretion}.  Note that, just like the accretion rate computed in Section \ref{sec:numerics:diagnostics:bh_growth}, the rate (\ref{mdot_flux}) represents a rate as measured by an observer at a large distance from the neutron star, i.e.~at $r \gg R$.  Comparing this rate with the rates computed in Section \ref{sec:estimates}, which represented those measured by a ``local asymptotic" observer at $\mbh \ll r \ll R$, again requires this local observer's lapse function $\alpha_\asym$ (see also Appendix \ref{app:schwarzschild}).  

Also note that the accretion rate discussed in \ref{sec:numerics:diagnostics:bh_growth} measures changes in the black hole's gravitational mass, while the flux (\ref{mdot_flux}) measures the accretion of rest mass.  In our numerical simulations we find that the black hole's gravitational mass grows at a rate somewhat larger than the rate of rest-mass accretion, which is presumably because the former includes the accretion of other forms of energy (in particular, internal thermal energy) in addition to rest-mass energy.

\section{Results}
\label{sec:results}

In this Section we compare our numerical results in Section \ref{sec:numerics} for our fiducial neutron star model hosting black holes with a wide range of different masses to our analytical estimates in Section \ref{sec:estimates}.

\subsection{Comparison with Bondi flow}
\label{sec:results:profiles}

We start with a comparison of fluid flow profiles.  In our numerical simulations, we focus on data in the vicinity of the black hole, at sufficiently late times so that the fluid has had enough time to settle down into steady-state accretion.  We compare these numerical results with those resulting from a direct integration of the ``relativistic Bondi-equations", i.e.~the equations describing spherically symmetric, steady-state,
adiabatic fluid flow in a Schwarzschild spacetime (see, e.g., Appendix G in ST and \cite{RicBS21a}).  

\begin{figure}[t]
    \centering
    \includegraphics[width = 0.45 \textwidth]{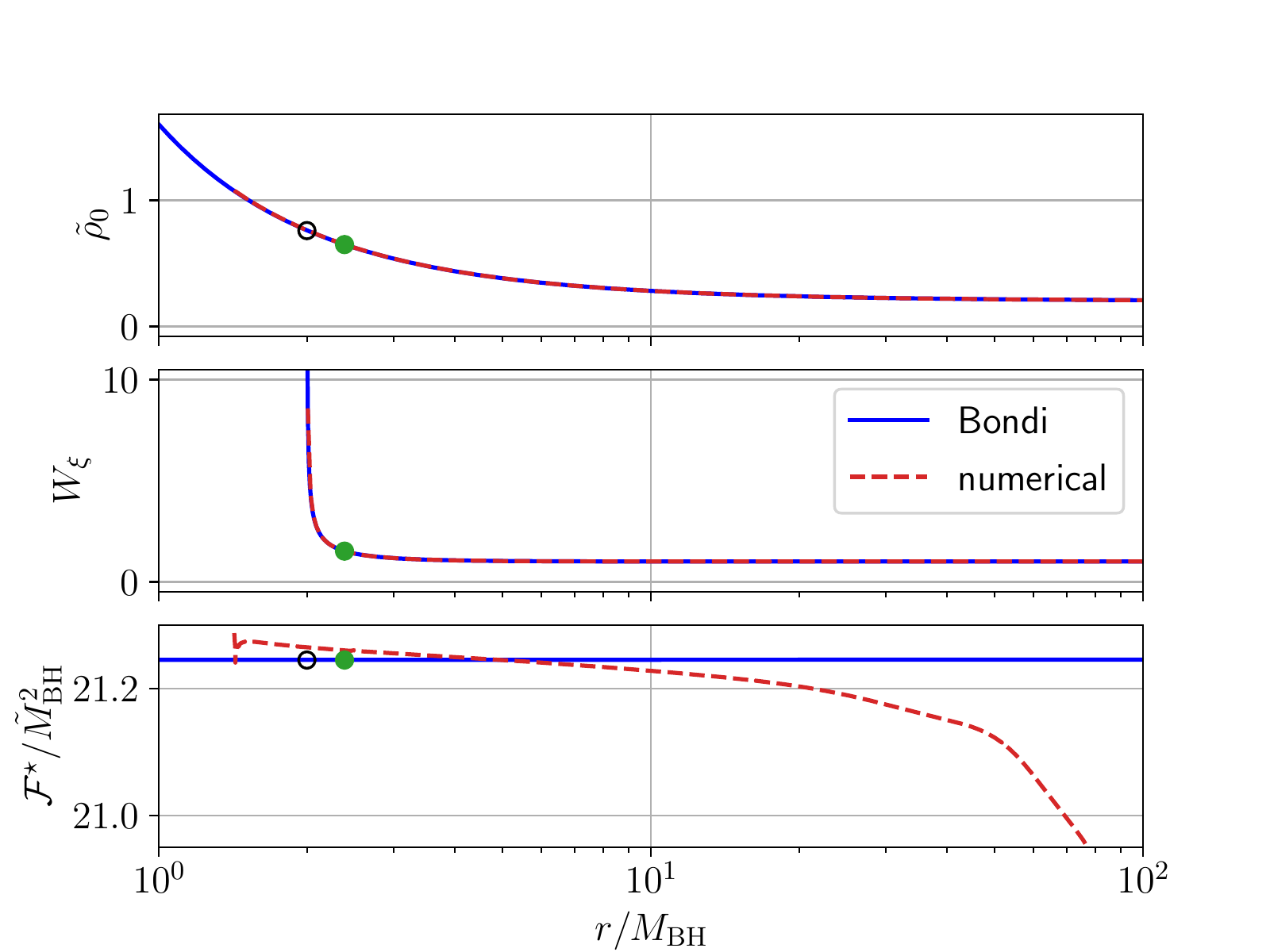} 
    \caption{Comparison of our numerical fluid flow profiles (dashed red lines) with those obtained from integrating the Bondi equations (solid blue lines), for our fiducial neutron star model hosting a black hole of initial mass $\mbhinittilde = 1.267 \times 10^{-6}$, at time $t = 1.59 \times 10^3 \mbh$ (see text for details).  The open black circles denote the location of the black hole horizon at $r = 2 \mbh$, where $r$ is the areal radius, while the solid green dots mark the location of the critical radius (which, in the Newtonian limit, reduces to the location at which the fluid flow becomes supersonic).  
    }
    \label{fig:bondi_compare}
\end{figure}

Since the coordinates used in our code are different from the Schwarzschild coordinates used in the usual construction of the steady-state Bondi solution, only scalar quantities -- for example the rest-mass density $\rho_0$ -- can be easily compared directly.  In order to compare an invariant measure of the fluid four-velocity $u^a$, we compute the ``gamma-factor" between an observer comoving with the fluid and a ``Killing observer", i.e.~a static observer whose four-velocity is aligned with a timelike Killing-vector $\xi^a = \partial/\partial t$,
\begin{equation} \label{W_xi}
    W_\xi = - \frac{\xi^a u_a}{( - \xi^a \xi_a)^{1/2}} = - \frac{u_t}{(-g_{tt})^{1/2}}.
\end{equation}
In Schwarzschild coordinates, this can be expressed as 
\begin{equation}
    \begin{split}
    W_\xi & = \alpha_{\rm S} u^{t_{\rm S}} = \left(1 - \frac{2\mbh}{r_{\rm S}} \right)^{1/2} u^{t_{\rm S}} \\
    & = \left\{ 1 + \left(1 - \frac{2 \mbh}{r_{\rm S}} \right)^{-1} (u^{r_{\rm S}})^2 \right\}^{1/2},
    \end{split}
\end{equation}
while, in our code, we evaluate 
\begin{equation}
    W_\xi = \frac{W ( \alpha - \beta_r v^r )}{\left( \alpha^2 - \beta_r \beta^r \right)^{1/2}}
\end{equation}
with $W=\alpha u^t$ and
\begin{equation}
v^r \equiv \frac{1}{W} \gamma^r_{a} u^a = \frac{u^r}{W} + \frac{\beta^r}{\alpha},
\end{equation}
where $v^i$ is the spatial projection of the four-velocity $u^a$, divided by $W$.  Here we assume not only that the spacetime is indeed approximately static, but also that the lapse $\alpha$ and the shift $\beta^r$ render this spacetime in a coordinate system that leaves the metric quantities nearly time-independent.
Note also that static observers exist only outside the black hole, so that we can evaluate $W_\xi$ only for $r \gtrsim 2 \mbh$. 

Finally, we can compare the flux (\ref{flux}) computed in our code with the accretion rate (\ref{mdot_bondi}) as predicted from the Bondi solution.

We show such a comparison for $\tilde \mpunc = 10^{-6}$ at (coordinate) time $t = 1.59 \times 10^3 \mbh$ in Fig.~\ref{fig:bondi_compare}.  For the rest-mass density $\rho_0$ and the gamma-factor $W_\xi$, the curves agree so well that they can hardly be distinguished in the figure.  The graphs for the accretion appear to differ more, but, at least in part, that is because the scale of the $y$-axis spans a much smaller range for the (almost) constant functions displayed in this panel.  In fact, even the accretion rates agree to within a small fraction of a percent in the vicinity of the black hole, demonstrating that the accretion onto an endoparasitic black hole inside a neutron star is well described by relativistic Bondi accretion. 

\subsection{Final Fate: Total Consumption}
\label{sec:results:complete}

For small initial black hole masses, the total accretion times are far too long for us to simulate the consumption of the entire neutron star numerically.  We have therefore performed simulations of such a complete consumption only for sufficiently large initial black hole masses.

\begin{figure}[t]
    \centering
    \includegraphics[width = 0.45 \textwidth]{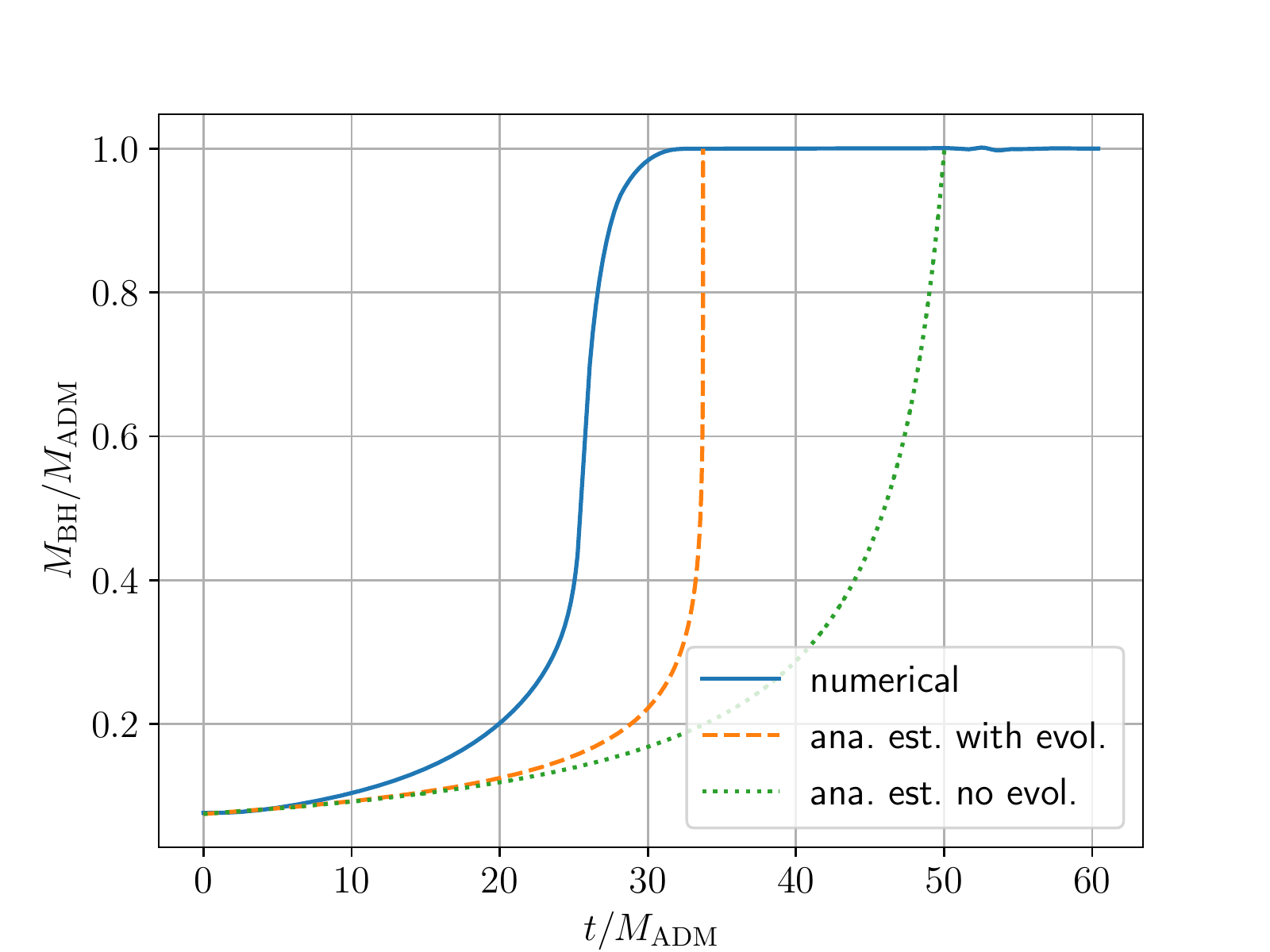} 
    \caption{The black hole mass $\mbh$ as a function of coordinate time $t$ in a simulation leading to complete collapse.   The solid line represents the numerical values of the black hole mass $\mbh$, starting with the initial black hole mass $\mbhinittilde = 0.0126$ and $\mbhinit/M_{\rm ADM} = 0.0761$.  At late times, after the black hole has consumed the entire neutron star, its mass agrees to high accuracy with the initial total gravitational mass $M_{\rm ADM}$.  The dashed line shows the analytical estimate (\ref{app_integral}) that takes into account effects of stellar evolution, while the dotted line shows the estimate (\ref{T_no_evolve_1}) that ignores these effects.  In this comparison we have adopted the value $\alpha_\asym \simeq 0.623$, even though for these simulations the black-hole mass is too large to identify a clean ``plateau" like the one shown in Fig.~\ref{fig:star_profile} (see text for details).
    }
    \label{fig:complete_collapse}
\end{figure}

As an example, we show in Fig.~\ref{fig:complete_collapse} results from a simulation with an initial black hole mass of $\mbhinittilde = 0.0126$.  These initial data have a total gravitational Arnowitt-Deser-Misner (ADM) mass of $\tilde M_{\rm ADM} = 0.1655$, so that $\mbhinit/M_{\rm ADM} = 0.0761$.  At early times, the black hole mass grows steadily, at a rate similar to those predicted by the analytical estimates of Section \ref{sec:estimates:bondi}.  At around $t \simeq 25 M_{\rm ADM}$, however, the consumption becomes dynamical, as described in Section \ref{sec:estimates:dynamical}.   Once the black hole has consumed the entire neutron star, its mass settles down to a value that agrees to within less than 0.1\% with the initial ADM mass of the system, which confirms the accuracy of our numerical simulations.

We also included in Fig.~\ref{fig:complete_collapse} the analytical estimates resulting from the integrations of Section \ref{sec:estimates:times}.  Specifically, the dashed line represents an estimate that includes the effects of stellar evolution (see Section \ref{sec:estimates:times:evolve}), while the dotted line ignores stellar evolution (Section \ref{sec:estimates:times:noevolve}).  

We caution that these comparisons should be considered qualitative for several reasons.  Perhaps most importantly, the initial ratio of the black hole to neutron star mass is too large for the asymptotic region in the core just beyond where the gas becomes bound to the black hole to be homogeneous, as required by the Bondi model. In addition, the final phase of accretion departs from secular to dynamical growth, and the Bondi rates we adopted throughout break down. Also, all the estimates of Section \ref{sec:estimates} are based on rates as observed by a ``local asymptotic", static observer.  Relating these ``local" rates to the ``global" rates computed in the numerical simulations requires the lapse function of a local observer, $\alpha_\asym$.  While this can be done rather well if $\mbh \ll \mns$, as shown in Fig.~\ref{fig:star_profile}, such a lapse function can no longer be identified unambiguously if $\mbh \lesssim \mns$. In fact, the entire notion of a local, static observer in a region with $\mbh \ll r \ll R$ no longer applies during the late stages. Finally, we used simple Newtonian arguments to model effects of stellar evolution (see Section \ref{sec:estimates:evolution}), which clearly do not apply in the late-time, dynamical collapse.  Were we able to follow the entire evolution for an initial black hole with $\mbh \ll \mns$ the number of decades in both time and increasing mass ratio during which the analytic curves closely match the numerical tracks in Fig~\ref{fig:complete_collapse} would be considerably larger. 

Despite all these disclaimers, we observe in Fig.~\ref{fig:complete_collapse} that the qualitative agreement between the numerical results and analytical estimates is reasonable. In particular, we see that including the effects of stellar evolution does improve this agreement.  In these comparisons, we adopted the value $\alpha_\asym = 0.623$ to convert from our analytic local to our numerical global observers. This was the value that we had identified for smaller black hole masses for this fiducial neutron star. But as we discussed above, in reality this value is no longer well defined at late times.

\subsection{Accretion rates}
\label{sec:results:rates}

\begin{table*}[t]
    \centering
    \begin{tabular}{c|c|c|c|c|c|c}
         $\tilde \mpunc$\footnote{Black hole puncture mass $\tilde \mpunc = K^{-1/2} \mpunc$ in our initial data; see Section \ref{sec:numerics:indata}.} & 
         $\mbhinittilde$\footnote{Initial irreducible mass $\mbhinittilde = K^{-1/2} \mbhinit$ of the black hole; see Section \ref{sec:numerics:diagnostics:mass}.} & 
         $\mbhinit/M_0$\footnote{Ratio between $\mbhinit$ and the neutron star rest mass $M_0$.} & 
         $\alpha_\asym$\footnote{Lapse of a ``local asymptotic" static observer; see, e.g., Fig.~\ref{fig:star_profile} for an example.}& 
         $\dotmbh / \alpha_\asym$\footnote{Mass-energy accretion rate from measurements of $\mbh$; see Section \ref{sec:numerics:diagnostics:bh_growth}.} & 
         $\flux(r_{\rm AH}) / \alpha_\asym$\footnote{Rest-mass accretion rate from flux across horizon; see Section \ref{sec:numerics:diagnostics:flux}.} & 
         $\dotmbhasym$\footnote{Rest-mass accretion rate from Bondi expressions; see Eq.~(\ref{mdot_fidu}).} 
         \\
         \hline 
         ~$10^{-3}$~ & ~$1.267 \times 10^{-3}$~ & ~$7.12 \times 10^{-3}$~ 
         & ~$0.616$~ & ~$4.58 \times 10^{-5}$~ & ~$3.44 \times 10^{-5}$~ & ~$3.41 \times 10^{-5}$~ 
         \\ 
         $10^{-4}$ & $1.267 \times 10^{-4}$ & $7.12 \times 10^{-4}$ 
         & $0.619$ & $4.67 \times 10^{-7}$ & $3.42 \times 10^{-7}$ & $3.41 \times 10^{-7}$  
         \\
         $10^{-5}$ & $1.267 \times 10^{-5}$ & $7.12 \times 10^{-5}$ 
         & $0.622$ & -- & $3.41 \times 10^{-9}$ & $3.41 \times 10^{-9}$ 
         \\
         $10^{-6}$ & $1.267 \times 10^{-6}$ & $7.12 \times 10^{-6}$ 
         & $0.623$ & -- & $3.40 \times 10^{-11}$ & $3.41 \times 10^{-11}$ 
         \\
         $10^{-7}$ & $1.267 \times 10^{-7}$ & $7.12 \times 10^{-7}$ 
         & $0.623$ & -- & $3.42 \times 10^{-13}$ & $3.41 \times 10^{-13}$ 
         \\
         $10^{-8}$ & $1.267 \times 10^{-8}$ & $7.12 \times 10^{-8}$ 
         & $0.623$ & -- & $3.43 \times 10^{-15}$ & $3.41 \times 10^{-15}$ 
         \\
         $10^{-9}$ & $1.267 \times 10^{-9}$ & $7.12 \times 10^{-9}$ 
         & $0.623$ & -- & $3.43 \times 10^{-17}$ & $3.41 \times 10^{-17}$ 
         \\
         $10^{-10}$ & $1.267 \times 10^{-10}$ & $7.12 \times 10^{-10}$ 
         & $0.623$ & -- & $3.43 \times 10^{-19}$ & $3.41 \times 10^{-19}$ 
    \end{tabular}
    \caption{
    Accretion rates for different black hole masses embedded in our fiducial neutron star model (see Table \ref{tab:ns}).  The accretion rates $\dotmbh$ and $\flux(r_{\rm AH})$ represent rates as measured by a static observer at infinity.  To compare these rates with those measured by a ``local asymptotic" static observer in the neutron star core, $\dotmbhasym$, we divide the former by the lapse $\alpha_\asym$ of the local observer.  The numerically measured rest-mass flux $\flux(r_{\rm AH})/\alpha_\asym$ (see Section \ref{sec:numerics:diagnostics:flux}) agrees very well with the analytical value $\dotmbhasym$ given by Bondi accretion (see \eq~\ref{mdot_fidu}).  Measuring the growth of the black hole horizon $\dotmbh/\alpha_\asym$ (see Section \ref{sec:numerics:diagnostics:bh_growth}) results in somewhat larger values, presumably because it includes internal thermal energy in addition to rest-mass energy.  Results in this table are also shown in Fig.~\ref{fig:accretion_rates}.}
    \label{tab:accretion}
\end{table*}

We summarize in Table \ref{tab:accretion} results from our simulations for a large range of initial black hole masses, spanning seven orders of magnitude in $\mbh / \mns$.  For each black hole mass, we compute in our code the accretion rates $\dotmbh$, directly from the growth of the black hole mass (for sufficiently large initial black holes) as discussed in Section \ref{sec:numerics:diagnostics:bh_growth}, and/or from the fluid flux, $\dotmbh = \flux(r_{\rm AH})$, as discussed in Section \ref{sec:numerics:diagnostics:flux}.  We also identify the lapse $\alpha_\asym$ of a ``local asymptotic" observer far inside the neutron star (see Fig.~\ref{fig:star_profile} for an example), and then divide the above ``global" rates by this lapse in order to compute the rates that such a local observer would measure.  The latter can also be estimated from the Bondi expression (\ref{mdot_fidu}), the result of which we list in the last column of Table \ref{tab:accretion}.  The entries in this Table are also shown in Fig.~\ref{fig:accretion_rates}.  

We clarify that, for the small black hole masses listed in Table \ref{tab:accretion}, we do not track the evolution to completion, until the entire star has been consumed, since doing so would require following the star for {\em many} more dynamical timescales than is computationally feasible.  Instead we evolve the system for a coordinate time of approximately $10^3 \mbh$, by which time the accretion has settled down into equilibrium and we can accurately measure the accretion rate. 

Most importantly we observe that the different measures of the accretion rates agree well with each other.  In our numerical simulations of Section \ref{sec:numerics}, the flux of (baryon) rest mass across the horizon agrees very well with the analytical accretion rates computed from relativistic Bondi accretion in Section \ref{sec:estimates}.  Determining the accretion rate from the growth of the black hole horizon, which provides a measure of the increase in total gravitational mass, results in a somewhat larger value, presumably because this includes internal thermal energy in addition to rest-mass energy.  Note, however, that this observation holds only for early times, while the accretion is described by Bondi accretion, and not for the last dynamical phase, during which most of the mass is accreted (see Fig.~\ref{fig:complete_collapse}).

Finally, we note that the initial black hole mass $\mbhinittilde$ agrees very well with the analytical estimate (\ref{m_irr_approx}) with $\psi_c = 1.27$ for our fiducial neutron star model (see Table \ref{tab:ns}).

\section{Discussion}
\label{sec:discussion}

We study in detail the process by which a small ``endoparasitic" black hole, residing at the center of a neutron star, consumes its host.  While a number of aspects of this problem have been studied before (see, e.g., \cite{CapPT13,EasL19,GenST20}), we expand on these treatments in a number of ways.

Building on our previous study of Bondi accretion for stiff EOSs \cite{RicBS21a,BauS21} we develop a quantitative analytical description of this accretion process.  In particular, this allows us to determine the constant of proportionality in the relation $\dotmbh \propto \mbh^2$ that some previous authors had adopted. We use these results to construct an approximate analytic model that tracks the secular evolution of the system as the black hole, assumed initially small, grows by accretion and ultimately consumes the entire neutron star.

We also perform numerical simulations of this accretion process, extending previous simulations (see \cite{EasL19}) to significantly smaller ratios $\mbh / \mns$ and simulating long enough for the
systems to achieve quasistationary accretion. Our numerical code adopts spherical polar coordinates (see \cite{BauMCM13,BauMM15}) with a logarithmic radial coordinate, which allows us to adequately resolve the vastly different length scales of the black hole and neutron star.   

As shown in Fig.~\ref{fig:accretion_rates}, our numerical results for the accretion rates agree remarkably well with those computed from relativistic Bondi accretion over many orders of magnitude in $\mbh / \mns$.  This establishes that the accretion onto small black holes at the center of neutron stars is indeed governed by secular Bondi accretion, and that the lifetimes of such stars are determined by these accretion rates.  In particular, this supports our finding reported in \cite{BauS21} that this lifetime is close to a nearly universal maximum lifetime that is roughly independent of the properties of the neutron star and its EOS, and depends on the initial black hole mass $\mbh$ only.

As an important application, our results corroborate arguments that use the current existence of neutron star populations to constrain either the contribution of primordial black holes to the dark matter content of the Universe, or that of dark matter particles that may form black holes at the center of neutron stars after they have been captured (see, e.g., \cite{GolN89,deLF10,BraL14,BraE15,CapPT13,BraLT18,EasL19,GenST20}).  These constraints are based on the notion that, given certain cosmological densities of these dark matter constituents and their masses, neutron stars would capture these objects and would then be consumed by the black holes after times that are in conflict with the ages of old neutron star populations. In particular, these arguments have been used to constrain the contribution of PBHs in the mass range $10^{-15} M_\odot \lesssim \mbh \lesssim 10^{-9} M_\odot$ (see, e.g., \cite{CapPT13,KueF17}).

\acknowledgments

We would like to thank Gordon Baym, Milton Ruiz, Maria Perez Mendoza, and Antonios Tsokaros for helpful conversations. CBR acknowledges support through an undergraduate research fellowship at Bowdoin College. This work was supported in parts by National Science Foundation (NSF) grants PHY-1707526 and PHY-2010394 to Bowdoin College, and NSF grants  PHY-1662211 and PHY-2006066 and National Aeronautics and Space Administration (NASA) grant 80NSSC17K0070 to the University of Illinois at Urbana-Champaign.

\begin{appendix}

\section{Black hole metric and Bondi accretion inside a star}
\label{app:schwarzschild}

In this Appendix we generalize the Schwarzschild metric so that it allows for an asymptotic region that is different from the Minkowski metric.  We then adopt this form of the metric to describe approximately a small black hole residing at the center of the neutron star, and to rederive the equations governing Bondi flow in this context.

\subsection{A general form of the Schwarzschild metric}

Following Section 5.1 in \cite{Car04}, we write the spherically symmetric, time-independent metric in vacuum  as
\begin{equation} \label{gen_metric}
ds^2 = - e^{2 A} dt^2 + e^{2 B} d\rareal^2 + \rareal^2 d\Omega^2,
\end{equation}
where, in this Appendix only, $\rareal$ is the areal radius (see 5.11 in \cite{Car04}, hereafter C5.11, except that we use $A$ and $B$ instead of $\alpha$ and $\beta$ in order to avoid confusion with the lapse function and the shift vector).  We first evaluate the combination $e^{2(B - A)} R_{tt} + R_{\rareal\rareal}$, where $R_{ab}$ is the Ricci tensor, in Einstein's vacuum field equations.  This yields $\partial_\rareal A + \partial_\rareal B = 0$ (see C5.16) and hence
\begin{equation} \label{AB}
    A = - B + C,
\end{equation}
where $C$ is a constant of integration.  We {\em usually} set this constant to zero (see C5.17), which results in $t$ being the proper time of a static observer at $\rareal \rightarrow \infty$, but here we will allow $C$ to remain non-zero, and will determine its value in Section \ref{sec:app:constants} below.

We next evaluate $R_{\theta\theta}$ in Einstein's equations, which now takes the form
\begin{equation}
    e^{2A} \left( 2 \rareal \, \partial_\rareal A + 1 \right) = e^{2C}
\end{equation}
(compare C5.18) and is solved by 
\begin{equation} \label{e2A}
    e^{2A} = e^{2C} - \frac{\kappa}{\rareal},
\end{equation}
where $\kappa$ is another constant of integration.  We {\em usually} identify this constant with $2M$, where $M$ is the gravitational mass of the black hole, but we will again postpone determining this constant until Section \ref{sec:app:constants}.

Using (\ref{e2A}) and (\ref{AB}) in (\ref{gen_metric}) we now find 
\begin{equation} \label{gen_metric2}
    ds^2 = -\left(e^{2C} - \frac{\kappa}{\rareal} \right) dt^2
    + \left( 1 - \frac{e^{-2C} \kappa}{\rareal} \right)^{-1} d \rareal^2
    + \rareal^2 d\Omega^2
\end{equation}
for the general form of the Schwarzschild metric.  Evidently, we recover the asymptotic-Minkowski form of the metric for $C = 0$ and $\kappa = 2 M$.

\subsection{Identification of constants}
\label{sec:app:constants}

We next identify the constants $C$ and $\kappa$ in the metric (\ref{gen_metric2}) by matching to our initial data describing a black hole embedded inside a neutron star.  We assume that the black hole's mass is small compared to the mass of the star, $\mbh \ll \mns$, and will restrict our analysis to a region that is large compared to $\mbh$, but small compared to $\mns$ (and the radius of the neutron star).  We may then approximate the neutron star's conformal factor as constant, $\psi_{\rm NS} \simeq \psi_c$ (see Table \ref{tab:ns} and the discussion in Section \ref{sec:numerics:indata}), and may neglect $u$ in (\ref{psi_ansatz}) (see Appendix \ref{sec:app:approx}).  We may also neglect the neutron star's contribution to the total mass-energy in this region, so that the metric (\ref{gen_metric2}) still provides an approximate solution to Einstein's equations, even though it was derived in vacuum. 

Recall that, in Section \ref{sec:numerics:indata}, we construct the initial spatial line element from
\begin{equation} \label{metric_iso}
    dl^2 = \psi^4 (d\riso^2 + \riso^2 d\Omega^2),
\end{equation}
where $\riso$ is the isotropic radius, and, by our assumption above, the conformal factor (\ref{psi_ansatz}) reduces to
\begin{equation} \label{psi_app}
    \psi = \psi_c + \frac{\mpunc}{2 \riso}.
\end{equation}
Identifying (\ref{metric_iso}) with the spatial part of (\ref{gen_metric2}), we obtain
\begin{equation} \label{psi}
    \riso \psi^2 = R
\end{equation}
and 
\begin{equation}
    \left( 1 - \frac{e^{-2C} \kappa}{\rareal} \right)^{-1/2} d\rareal = \psi^2 d\riso = \rareal \frac{dr}{r}
\end{equation}
or
\begin{equation}
    \frac{d\rareal}{\rareal^{1/2} (R - e^{-2C} \kappa)^{1/2}} = \frac{dr}{r}.
\end{equation}
Integration of both sides yields
\begin{equation}
    D \riso = 2 \left( \rareal + \sqrt{\rareal^2 - \rareal e^{-2C} \kappa} - \frac{e^{-2C}\kappa}{2} \right)  
\end{equation}
where $D$ is a constant of integration, which we can solve for $\rareal$ to find
\begin{equation}
    \rareal = \frac{D \riso}{4} \left( 1 + \frac{e^{-2C} \kappa}{D \riso} \right)^2. 
\end{equation}
We next use (\ref{psi}) to write
\begin{equation}
    \psi = \left( \frac{\rareal}{\riso} \right)^{1/2}
    = \frac{D^{1/2}}{2} \left( 1 + \frac{e^{-2C} \kappa}{D \riso} \right)
\end{equation}
and compare this with (\ref{psi_app}) to identify
\begin{equation}
    D = 4 \psi_c^2
\end{equation}
and
\begin{equation}
    e^{-2C} \kappa = 2 \psi_c \mpunc = 2 M_{\rm irr},
\end{equation}
where we have used (\ref{m_irr_approx}), derived under the same assumptions as our treatment here, in the last equality.

We now evaluate the time part of the metric (\ref{gen_metric2}).  In particular, consider a ``static asymptotic" observer as introduced in Section \ref{sec:estimates:bondi}, i.e.~an observer at areal radius $\rareal \gg M_{\rm irr}$, but far inside the neutron star so that our assumptions above apply.  According to (\ref{gen_metric2}), the proper time of such an observer advances at a rate $d\tau_\asym = e^C dt$, meaning that we may identify $e^C$ with the lapse function of such an observer, $\alpha_\asym$.

Finally, we can insert our results into (\ref{gen_metric2}) to find the metric describing the spacetime in the vicinity of a black hole embedded in a neutron star,
\begin{align} \label{gen_metric3}
    ds^2  = & - \alpha_\asym^2 \left(1 - \frac{2 \mbh}{\rareal} \right) dt^2 +  \left(1 - \frac{2 \mbh}{\rareal} \right)^{-1} d\rareal^2 \nonumber \\
    &  + \rareal^2 d\Omega^2,
\end{align}
where we have identified the black hole mass with its irreducible mass, $\mbh = M_{\rm irr}$.

\longcomment{
{\red [STU: take a look at this...] As an aside, it may also be instructive to compare the above with the analytical expressions for a static, spherical shell surrounding a black hole, see \cite{FraHK90}.  The spatial part of their interior metric agrees with that of (\ref{gen_metric3}), while, following \cite{FraHK90}, the proper time of a static observer inside the shell (but outside the black hole) advances at a rate
\begin{equation}
d\tau = \left( 1 - \frac{2 \mbh}{\rareal} \right)^{1/2} d\bar t
= \frac{\alpha_{M_{\rm ADM}}}{\alpha_{M_{\rm BH}}} \left( 1 - \frac{2 \mbh}{\rareal} \right)^{1/2} d t.
\end{equation}
Here we have defined $\alpha_{M_{\rm ADM}} \equiv (1 - 2 M_{\rm ADM}/R_{\rm sh})^{1/2}$ and $\alpha_{M_{\rm BH}} = (1 - 2 \mbh/ R_{\rm sh})^{1/2}$, $R_{\rm sh}$ is the proper radius of the shell, and $M_{\rm ADM}$ the total gravitational mass of the system.  We now identify
\begin{equation}
    \alpha_\asym = \frac{\alpha_{M_{\rm ADM}}}{\alpha_{M_{\rm BH}}}
\end{equation}
and see that the interior metric of \cite{FraHK90} agrees with our metric (\ref{gen_metric3}) above.  To leading order we may expand
\begin{equation}
    \alpha_\asym \simeq \left( 1 - \frac{2 (M_{\rm ADM} - M_{\rm BH}}{R_{\rm sh}} \right)^{1/2}, 
\end{equation}
so that we may loosely interpret $\alpha_\asym$ as the lapse function resulting form the shell (or star) alone, as measured far from the black hole, but inside the shell (or star).
}}

\subsection{Bondi accretion}

The equations governing Bondi accretion, i.e.~the continuity equation and the Euler equation of relativistic hydrodynamics for stationary, adiabatic, and spherically symmetric fluid flow, are usually derived assuming a Schwarzschild metric that asymptotes to a Minkowski metric.  While our metric (\ref{gen_metric3}) takes a slightly different form, both the continuity equation and the Euler equation take the exact same form.  That means that the results for relativistic Bondi accretion (as presented, for example, in Appendix G of ST or in \cite{RicBS21a}) can be adopted without change, provided that the black hole mass $\mbh$ is identified with its irreducible mass $M_{\rm irr}$.  In particular, the accretion rate is given by \eq~(\ref{mdot_bondi}).

There remains one ambiguity, however, namely the meaning of the time derivative in $\dotmbh$ in the accretion rate, i.e.~whose time we refer to in this derivative.  In order to clarify this, we compute the radial component of the fluid flux as observed by a static observer at an arbitrary radius $\rareal$ outside the black hole horizon (i.e.~not necessarily at $\rareal \gg \mbh$, but well inside the neutron star).  Specifically, we take the dot product between the fluid flux ${\bf J} = \rho_0 {\bf u}$ (where we use bold-face to denote a vector) with the orthonormal basis one-form $\tilde {\boldsymbol \omega}^{\hat \rareal} = (1 - 2 \mbh/\rareal)^{-1/2} \tilde {\boldsymbol \omega}^{\rareal}$ to find
\begin{equation}
    J^{\hat \rareal} = \tilde {\boldsymbol \omega}^{\hat \rareal} \cdot {\bf J} = \left( 1 - \frac{2 \mbh}{\rareal} \right)^{-1/2} J^\rareal.
\end{equation}
We now multiply with $- 4 \pi \rareal^2$ and define $u \equiv - u^{\rareal}$ (which is positive for inflowing matter) to obtain the accretion rate $dM/d\tau$ as measured by this observer,
\begin{equation}
    \frac{d\mbh}{d\tau} = - 4 \pi \rareal^2 J^{\hat \rareal} 
    = 4 \pi \rareal^2 \left( 1 - \frac{2 \mbh}{\rareal} \right)^{-1/2} \rho_0 u.
\end{equation}
Since, for this observer, $d\tau = \alpha_\asym (1 - 2 \mbh/\rareal)^{1/2} dt$, we may rewrite the above expression as
\begin{equation} \label{mdot_infty}
    \frac{d\mbh}{d t} = 4 \pi \alpha_\asym \rareal^2 \rho_0 u,
\end{equation}
which implies
\begin{equation} \label{mdot_locasym}
    \frac{d\mbh}{d \tau_\asym} 
    = 4 \pi \rareal^2 \rho_0 u,
\end{equation}
where $\tau_\asym$ is again the proper time of a static ``local asymptotic" observer.  We see that the ``usual" expression for the Bondi accretion rate (e.g.~14.3.18 or G.21 in ST) refers to a time derivative with respect to the time as observed by a static asymptotic observer, which, in our case, means a ``local asymptotic" observer -- far from the black hole, but well inside the neutron star.  In (\ref{mdot_bondi}) and elsewhere we emphasize this by denoting the time derivative with a star, $\dotmbhasym = d \mbh / d\tau_\asym$.  We also see that the accretion rates observed by a local asymptotic observer and an observer at infinity are related by
\begin{equation} \label{relate_rates_app}
    \dotmbhasym = \frac{d\mbh}{d \tau_\asym} = \frac{1}{\alpha_\asym} \, \frac{d\mbh}{dt} = \frac{1}{\alpha_\asym} \dotmbh,
\end{equation}
as we have observed already in (\ref{relate_rates}).



\section{Integration of \eq~(\ref{dydT})}
\label{sec:integral}

In this appendix we outline how the 
differential equation (\ref{dydT}) can be solved analytically.

We first separate variables to obtain
\begin{equation}
    d T = - \frac{y^{7/2} dy}{(y_0 - y)^2} 
\end{equation}
and then integrate to find
\begin{equation}
    T = - I,
\end{equation}
where we have assumed that the initial time is chosen to vanish, $T_i = 0$, and where $I$ is given by
\begin{equation} \label{integral1}
    I = \int \frac{y^{7/2} dy}{(y_0 - y)^2}.
\end{equation}
This integral can now be integrated as follows.  

Using partial fractions, we rewrite (\ref{integral1}) as
\begin{eqnarray}
I & = & \int \frac{y^{7/2} dy}{(y_0 - y)^2} \nonumber \\
& = & \int y^{3/2} \frac{y^2 + (y_0 - y)^2 - (y_0 - y)^2}{(y_0 - y)^2} dy 
\nonumber \\
& = & \int y^{3/2} \frac{(y_0 - y)^2 + 2 y y_0 - y_0^2}{(y_0 - y)^2} dy \nonumber \\
& = & \int y^{3/2} dy + 2 y_0 \int y^{3/2} \frac{y - y_0 / 2}{(y_0 - y)^2} dy .
\end{eqnarray}
Repeating the process twice more, we obtain
\begin{eqnarray}
I & = &  \int y^{3/2} dy + 2 y_0 \int y^{1/2} dy + 3 y_0^2 \int y^{-1/2} dy \nonumber \\
& & +  4 y_0^3 \int y^{-1/2} \frac{y - 3 y_0 / 4}{(y_0 - y)^2} dy. \label{longint}
\end{eqnarray}
We now split the last integral into two terms,
\begin{eqnarray}
& & y_0^3 \int y^{-1/2} \frac{4 y - 3 y_0}{(y_0 - y)^2} dy \\
& & ~~~~ =  
4 y_0^3 \int \frac{y^{1/2} }{(y_0 - y)^2} dy 
 - 3 y_0^4 \int \frac{y^{-1/2}}{(y_0 - y)^2} dy \nonumber
\end{eqnarray}
and use a hyperbolic trig substitution
\begin{equation} \label{tanhsub}
y = y_0 \tanh^2 x
\end{equation}
in both integrals, resulting in
\begin{equation}
4 y_0^3 \int \frac{y^{1/2} }{(y_0 - y)^2} dy = 8 y_0^{5/2} \int \sinh^2 x \, dx
\end{equation}
and
\begin{equation}
3 y_0^4 \int \frac{y^{-1/2}}{(y_0 - y)^2} dy = 6 y_0^{5/2} \int \cosh^2 x \, dx.
\end{equation}
Since
\begin{equation}
 \int \sinh^2 x \, dx = \frac{1}{2} \sinh x \cosh x  - \frac{x}{2}
\end{equation}
and
\begin{equation}
 \int \cosh^2 x \, dx = \frac{1}{2} \sinh x \cosh x  + \frac{ x }{2},
\end{equation}
as can be seen using integration by parts, we can
combine results to find
\begin{equation} 
y_0^3 \int y^{-1/2} \frac{4 y - 3 y_0}{(y_0 - y)^2} dy  = y_0^{5/2} \left( \sinh x \cosh x - 7  x  \right). 
\end{equation}
We now rewrite $\sinh x \cosh x$ in terms of $\tanh x$
and insert the substitution (\ref{tanhsub}) to obtain
\begin{eqnarray}
& & y_0^3 \int y^{-1/2} \frac{4 y - 3 y_0}{(y_0 - y)^2} dy  \\
& & ~~~~ = y_0^{3} \frac{y^{1/2}}{y_0 - y} 
 - 7 y_0^{5/2} \tanh^{-1} (y / u_0)^{1/2} .
\nonumber
\end{eqnarray} 
Finally we insert this expression into (\ref{longint}), and carry out the remaining integrations to find
\begin{eqnarray}
I & = & \frac{2}{5} y^{5/2}  + \frac{4}{3} y_0 y^{3/2}  + 6 y_0^2  y^{1/2} \\ 
& & + y_0^{3}  \frac{y^{1/2}}{y_0 - y}  - 7 y_0^{5/2}  \tanh^{-1} (y / y_0)^{1/2}.
\nonumber 
\end{eqnarray}
Combining the first four terms and using 
\begin{equation}
    \tanh^{-1} (x) = \frac{1}{2} \ln \left( \frac{1 + x}{1 - x} \right)
\end{equation}
we can also write this result as
\begin{eqnarray} \label{app_integral}
I & = & y^{1/2} \frac{6y^3 + 14 y_0y^2 + 70 y_0^2 y - 105 y_0^3}{15(y - y_0)} \nonumber \\ 
& & - \frac{7}{2} y_0^{5/2} \ln \left( \frac{ y_0^{1/2} + y^{1/2}}{y_0^{1/2} - y^{1/2}} \right).
\end{eqnarray}
Recall that $1 - y = (\mbh - \mbhinit)/\mnsinit$ measures the fractional increase in the black hole mass (see \eqs~\ref{mass_conserv} and \ref{yfac}), and that $T = - I$ is proportional to the time as measured by a local asymptotic static observer (see \eq~\ref{Tfac}).

\section{An Approximate Analytical Solution to the Hamiltonian constraint}
\label{sec:app:approx}

In this Appendix we present an approximate but analytical solution to the Hamiltonian constraint (\ref{Ham_u}),
\begin{equation} \label{Ham_app}
\bar D^2 u = - 2 \pi \left\{ \left( \psi_{\rm NS} + \psi_{\rm BH} + u \right)^{5 + n} - \psi_{\rm NS}^{5 + n} \right\} \bar \rho.
\end{equation}
We have solved this equation to high precision in Section \ref{sec:numerics:indata}, and our goal here is not to reproduce that solution quantitatively; instead, we adopt some simple arguments that allow us to understand the qualitative behavior of the solution $u$.  In particular, we will find that $u$ approaches zero {\em everywhere} as $\mbh/\mns \rightarrow 0$, which justifies neglecting $u$, for example, in (\ref{psi_drop_u}) and (\ref{psi_app}).

In fact, we start by assuming that $u \ll \psi_{\rm NS}$, so that we may neglect this term on the right-hand side of (\ref{Ham_app}), ultimately verifying the validity of this step by showing that it is well satisfied by our final solution.   We further approximate the neutron-star conformal factor $\psi_{\rm NS}$ by using its central value throughout the entire interior of the star, i.e.~for all $r < R$, and we replace the density $\bar \rho$ by its average value $\bar \rho_{\rm ave} = \bar \rho_c / \delta$, where we adopt the Newtonian central condensation $\delta = 3.29$ for a $\Gamma = 2$ polytrope (see Table \ref{tab:polytropes}).  While we will see that $u$ does indeed become arbitrarily small as $\mpunc$ approaches zero, the latter approximation is rather crude and introduces discrepancies, but only of order unity.

We next observe that, even in the interior of the star, the right-hand side of (\ref{Ham_app}) behaves differently depending on whether $\psi_{\rm BH}$ or $\psi_{\rm NS}$ is greater.  If $\psi_{\rm BH} = \mpunc / (2 r) \gg \psi_{\rm NS}$, and assuming $n < -5$, we have 
\begin{equation}
    \left( \psi_{\rm NS} + \psi_{\rm BH} \right)^{5 + n} - \psi_{\rm NS}^{5 + n} \simeq \psi_{\rm BH}^{5 + n} - \psi_{\rm NS}^{5 + n} \simeq - \psi_{\rm NS}^{5 + n}.
\end{equation}
On the other hand, if $\psi_{\rm BH} \ll \psi_{\rm NS}$, we may expand
\begin{equation}
    \left( \psi_{\rm NS} + \psi_{\rm BH} \right)^{5 + n} - \psi_{\rm NS}^{5 + n} \simeq 
    (5 + n) \psi_{\rm NS}^{4 + n} \psi_{\rm BH}.
\end{equation}
Recognizing that $\psi_{\rm BH} = \psi_{\rm NS}$ at the approximate location of the apparent horizon
\begin{equation}
    r_{\rm AH} = \frac{\mpunc}{2 \psi_{\rm NS}}
\end{equation}
identified in (\ref{r_AH}), we may approximate \eq~(\ref{Ham_app}) as
\begin{equation} \label{Ham_app2}
    \bar D^2 u = \left\{
    \begin{array}{ll}
         s & r < r_{\rm AH} \\
         \tilde s / r ~~~ &  r_{\rm AH} \leq r < R\\
         0 & r \geq R,
    \end{array}
    \right.
\end{equation}
where we have defined 
\begin{equation}
    s = 2 \pi \psi_{\rm NS}^{5+n} \bar \rho
\end{equation}
and
\begin{equation}
    \tilde s = - \pi (5 + n) \psi_{\rm NS}^{4 + n} \mpunc \, \bar \rho = - (5 + n) \, s \, r_{\rm AH}.
\end{equation}
Since $\bar D^2$ is the flat Laplace operator in spherical symmetry,
\begin{equation}
    \bar D^2 u = \frac{1}{r^2} \frac{d}{dr} \left( r^2 \frac{d u}{dr} \right),
\end{equation}
piece-wise solutions to (\ref{Ham_app2}) are given by
\begin{equation} \label{u_approx_sol}
    u = \left\{
    \begin{array}{ll}
         \displaystyle \frac{1}{6} s r^2 + C_{\rm I} 
         & r < r_{\rm AH} \\[3mm]
         \displaystyle \frac{1}{2} \tilde s r + C_{\rm II} + \frac{D_{\rm II}}{r} ~~~ &  r_{\rm AH} \leq r < R\\[3mm]
         \displaystyle \frac{C_{\rm III}}{r} & r \geq R.
    \end{array}
    \right.
\end{equation}
Here we have assumed regularity at the origin $r = 0$ and $u \rightarrow 0$ as $r \rightarrow \infty$, and the four constants $C_{\rm I}$, $C_{\rm II}$, $D_{\rm II}$, and $C_{\rm III}$ are constants of integration.  We can determine the latter by requiring that both $u$ and its first derivative are continuous at both $r_{\rm AH}$ and $R$. Evaluating these four conditions yields
\begin{equation} \label{constants}
    \begin{split}
        C_{\rm I} & = - \tilde s \,(R - r_{\rm AH}) - \frac{1}{2} s \, r_{\rm AH}^2 \\
        C_{\rm II} & = - \tilde s \, R \\
        D_{\rm II} & = \frac{1}{2} \tilde s \, r_{\rm AH}^2 - \frac{1}{3} s \, r_{\rm AH}^3 \\
        C_{\rm III} & = \frac{1}{2} \tilde s \, (r_{\rm AH}^2 - R^2) - \frac{1}{3} s \, r_{\rm AH}^3.
    \end{split}
\end{equation}

\begin{figure}[t]
    \centering
    \includegraphics[width = 0.45 \textwidth]{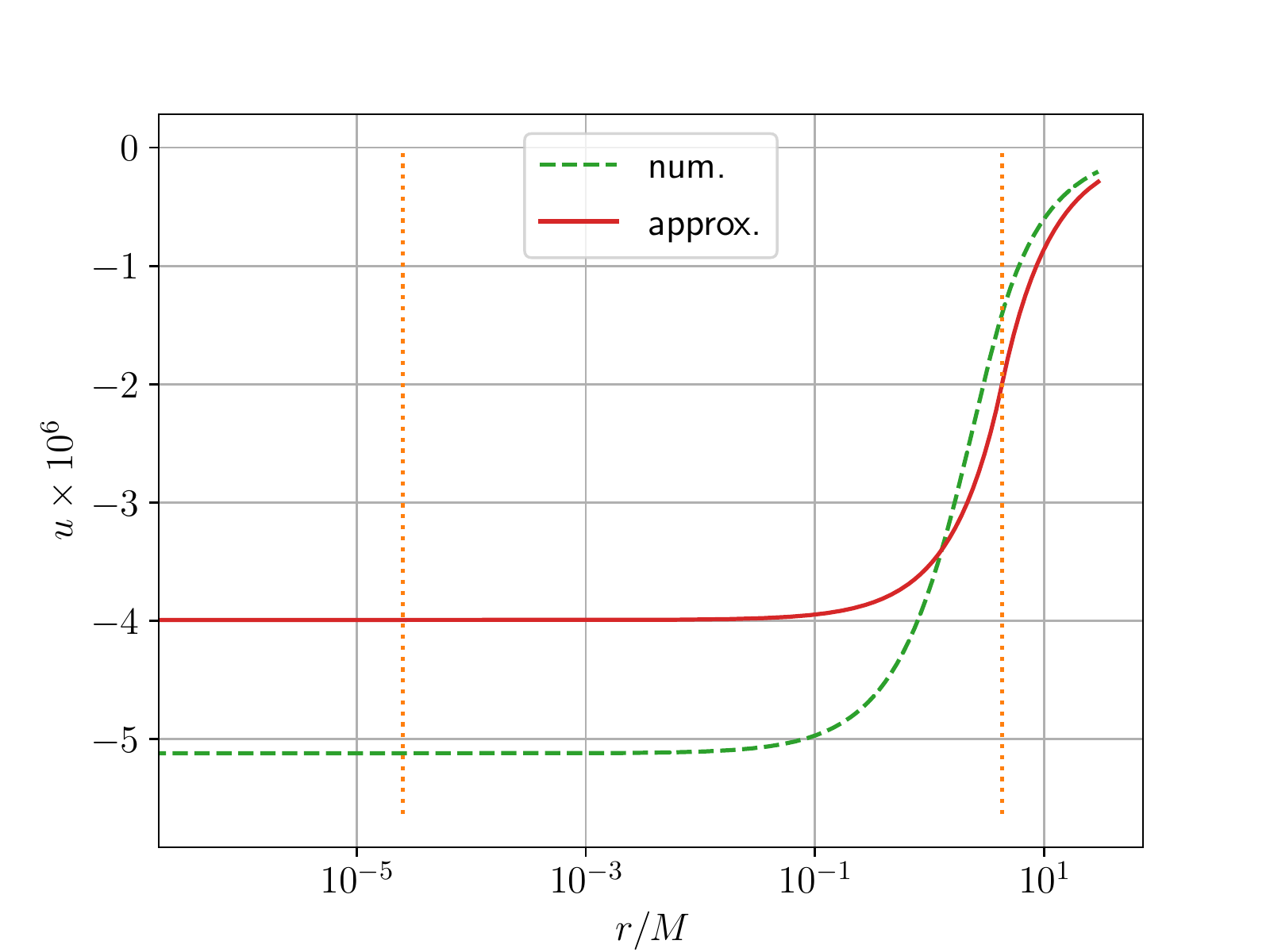} 
    \caption{The numerical solution $u$ for our fiducial neutron-star model and $\tilde \mpunc = 10^{-5}$ with $n = -6$ (dashed line), together with the approximate solution (\ref{u_approx_sol}) (solid line).  The two vertical dotted lines mark the approximate locations of the apparent horizon and the stellar surface.}
    \label{fig:u_approx}
\end{figure}

In Fig.~\ref{fig:u_approx} we compare the numerical solution for $u$, for our fiducial neutron-star model and $\tilde \mpunc = 10^{-5}$, with the approximate analytical solution (\ref{u_approx_sol}).  As expected, the two solutions do not agree quantitatively, but they nevertheless show very similar qualitative behavior.  Based on this qualitative agreement we may make the following observations about the properties of solutions $u$ to the Hamiltonian constraint (\ref{Ham_u}):
\begin{itemize}
    \item The largest value of $u$, in magnitude, occurs at the center, where it is dominated by the term 
    \begin{equation}
        u_c \simeq - \tilde s R \propto - \bar \rho \mpunc R. 
    \end{equation}
    We see that $u$ is proportional to $\mpunc$, and vanishes everywhere in the limit $\mpunc \rightarrow 0$.  This justifies our above approximation to neglect $u$ on the right-hand side of (\ref{Ham_app}), as well as in (\ref{psi_drop_u}) and (\ref{psi_app}).

    \item Similarly, $u$ is proportional to $\mpunc$ at $r \simeq r_{\rm AH}$, while $\psi_{\rm BH}$ is of order unity there.  This justifies our approximation in Section \ref{sec:numerics:diagnostics:mass} to neglect $u$ when computing an approximate location and area of the apparent horizon in the limit $\mpunc \rightarrow 0$.
    
    \item  In the exterior of the star, the contribution of $u$ to the conformal factor is proportional to 
    \begin{equation}
        u \simeq - \frac{\tilde s \, R^2}{2 r} 
        \sim - \frac{\rho R^2 \mpunc}{2r}. 
    \end{equation}
     We now assume that, in the far-field, we can write $\psi \simeq 1 + M_{\rm ADM} / (2r)$, and adopt a weak-field approximation $M_{\rm ADM} \simeq \mns + \mbh + U$, where $U$ is the interaction energy, i.e.~the potential energy resulting from having placed a black hole at the center of the neutron star.  We then have
    \begin{equation}
    \begin{array}{rcl}
        U & \simeq &  M_{\rm ADM} - \mns - \mbh \\ 
        & \simeq & 2r (\psi - \psi_{\rm NS} - \psi_{\rm BH}) \\
        & = & 2r u \sim - \rho R^2 \mpunc 
        \sim \displaystyle - \frac{\mns \mpunc}{R},
    \end{array}
    \end{equation}
    in accordance with the Newtonian expression for the interaction energy.
\end{itemize}

\end{appendix}


\begin{thebibliography}{51}%
\makeatletter
\providecommand \@ifxundefined [1]{%
 \@ifx{#1\undefined}
}%
\providecommand \@ifnum [1]{%
 \ifnum #1\expandafter \@firstoftwo
 \else \expandafter \@secondoftwo
 \fi
}%
\providecommand \@ifx [1]{%
 \ifx #1\expandafter \@firstoftwo
 \else \expandafter \@secondoftwo
 \fi
}%
\providecommand \natexlab [1]{#1}%
\providecommand \enquote  [1]{``#1''}%
\providecommand \bibnamefont  [1]{#1}%
\providecommand \bibfnamefont [1]{#1}%
\providecommand \citenamefont [1]{#1}%
\providecommand \href@noop [0]{\@secondoftwo}%
\providecommand \href [0]{\begingroup \@sanitize@url \@href}%
\providecommand \@href[1]{\@@startlink{#1}\@@href}%
\providecommand \@@href[1]{\endgroup#1\@@endlink}%
\providecommand \@sanitize@url [0]{\catcode `\\12\catcode `\$12\catcode
  `\&12\catcode `\#12\catcode `\^12\catcode `\_12\catcode `\%12\relax}%
\providecommand \@@startlink[1]{}%
\providecommand \@@endlink[0]{}%
\providecommand \url  [0]{\begingroup\@sanitize@url \@url }%
\providecommand \@url [1]{\endgroup\@href {#1}{\urlprefix }}%
\providecommand \urlprefix  [0]{URL }%
\providecommand \Eprint [0]{\href }%
\providecommand \doibase [0]{https://doi.org/}%
\providecommand \selectlanguage [0]{\@gobble}%
\providecommand \bibinfo  [0]{\@secondoftwo}%
\providecommand \bibfield  [0]{\@secondoftwo}%
\providecommand \translation [1]{[#1]}%
\providecommand \BibitemOpen [0]{}%
\providecommand \bibitemStop [0]{}%
\providecommand \bibitemNoStop [0]{.\EOS\space}%
\providecommand \EOS [0]{\spacefactor3000\relax}%
\providecommand \BibitemShut  [1]{\csname bibitem#1\endcsname}%
\let\auto@bib@innerbib\@empty
\bibitem [{\citenamefont {{Goldman}}\ and\ \citenamefont
  {{Nussinov}}(1989)}]{GolN89}%
  \BibitemOpen
  \bibfield  {author} {\bibinfo {author} {\bibfnamefont {I.}~\bibnamefont
  {{Goldman}}}\ and\ \bibinfo {author} {\bibfnamefont {S.}~\bibnamefont
  {{Nussinov}}},\ }\bibfield  {title} {\bibinfo {title} {{Weakly interacting
  massive particles and neutron stars}},\ }\href
  {https://doi.org/10.1103/PhysRevD.40.3221} {\bibfield  {journal} {\bibinfo
  {journal} {\prd}\ }\textbf {\bibinfo {volume} {40}},\ \bibinfo {pages} {3221}
  (\bibinfo {year} {1989})}\BibitemShut {NoStop}%
\bibitem [{\citenamefont {{de Lavallaz}}\ and\ \citenamefont
  {{Fairbairn}}(2010)}]{deLF10}%
  \BibitemOpen
  \bibfield  {author} {\bibinfo {author} {\bibfnamefont {A.}~\bibnamefont {{de
  Lavallaz}}}\ and\ \bibinfo {author} {\bibfnamefont {M.}~\bibnamefont
  {{Fairbairn}}},\ }\bibfield  {title} {\bibinfo {title} {{Neutron stars as
  dark matter probes}},\ }\href {https://doi.org/10.1103/PhysRevD.81.123521}
  {\bibfield  {journal} {\bibinfo  {journal} {\prd}\ }\textbf {\bibinfo
  {volume} {81}},\ \bibinfo {eid} {123521} (\bibinfo {year} {2010})},\ \Eprint
  {https://arxiv.org/abs/1004.0629} {arXiv:1004.0629 [astro-ph.GA]}
  \BibitemShut {NoStop}%
\bibitem [{\citenamefont {{Bramante}}\ and\ \citenamefont
  {{Linden}}(2014)}]{BraL14}%
  \BibitemOpen
  \bibfield  {author} {\bibinfo {author} {\bibfnamefont {J.}~\bibnamefont
  {{Bramante}}}\ and\ \bibinfo {author} {\bibfnamefont {T.}~\bibnamefont
  {{Linden}}},\ }\bibfield  {title} {\bibinfo {title} {{Detecting Dark Matter
  with Imploding Pulsars in the Galactic Center}},\ }\href
  {https://doi.org/10.1103/PhysRevLett.113.191301} {\bibfield  {journal}
  {\bibinfo  {journal} {\prl}\ }\textbf {\bibinfo {volume} {113}},\ \bibinfo
  {eid} {191301} (\bibinfo {year} {2014})},\ \Eprint
  {https://arxiv.org/abs/1405.1031} {arXiv:1405.1031 [astro-ph.HE]}
  \BibitemShut {NoStop}%
\bibitem [{\citenamefont {{Bramante}}\ and\ \citenamefont
  {{Elahi}}(2015)}]{BraE15}%
  \BibitemOpen
  \bibfield  {author} {\bibinfo {author} {\bibfnamefont {J.}~\bibnamefont
  {{Bramante}}}\ and\ \bibinfo {author} {\bibfnamefont {F.}~\bibnamefont
  {{Elahi}}},\ }\bibfield  {title} {\bibinfo {title} {{Higgs portals to pulsar
  collapse}},\ }\href {https://doi.org/10.1103/PhysRevD.91.115001} {\bibfield
  {journal} {\bibinfo  {journal} {\prd}\ }\textbf {\bibinfo {volume} {91}},\
  \bibinfo {eid} {115001} (\bibinfo {year} {2015})},\ \Eprint
  {https://arxiv.org/abs/1504.04019} {arXiv:1504.04019 [hep-ph]} \BibitemShut
  {NoStop}%
\bibitem [{\citenamefont {{Capela}}\ \emph {et~al.}(2013)\citenamefont
  {{Capela}}, \citenamefont {{Pshirkov}},\ and\ \citenamefont
  {{Tinyakov}}}]{CapPT13}%
  \BibitemOpen
  \bibfield  {author} {\bibinfo {author} {\bibfnamefont {F.}~\bibnamefont
  {{Capela}}}, \bibinfo {author} {\bibfnamefont {M.}~\bibnamefont
  {{Pshirkov}}},\ and\ \bibinfo {author} {\bibfnamefont {P.}~\bibnamefont
  {{Tinyakov}}},\ }\bibfield  {title} {\bibinfo {title} {{Constraints on
  primordial black holes as dark matter candidates from capture by neutron
  stars}},\ }\href {https://doi.org/10.1103/PhysRevD.87.123524} {\bibfield
  {journal} {\bibinfo  {journal} {\prd}\ }\textbf {\bibinfo {volume} {87}},\
  \bibinfo {eid} {123524} (\bibinfo {year} {2013})},\ \Eprint
  {https://arxiv.org/abs/1301.4984} {arXiv:1301.4984 [astro-ph.CO]}
  \BibitemShut {NoStop}%
\bibitem [{\citenamefont {{Bramante}}\ \emph {et~al.}(2018)\citenamefont
  {{Bramante}}, \citenamefont {{Linden}},\ and\ \citenamefont
  {{Tsai}}}]{BraLT18}%
  \BibitemOpen
  \bibfield  {author} {\bibinfo {author} {\bibfnamefont {J.}~\bibnamefont
  {{Bramante}}}, \bibinfo {author} {\bibfnamefont {T.}~\bibnamefont
  {{Linden}}},\ and\ \bibinfo {author} {\bibfnamefont {Y.-D.}\ \bibnamefont
  {{Tsai}}},\ }\bibfield  {title} {\bibinfo {title} {{Searching for dark matter
  with neutron star mergers and quiet kilonovae}},\ }\href
  {https://doi.org/10.1103/PhysRevD.97.055016} {\bibfield  {journal} {\bibinfo
  {journal} {\prd}\ }\textbf {\bibinfo {volume} {97}},\ \bibinfo {eid} {055016}
  (\bibinfo {year} {2018})},\ \Eprint {https://arxiv.org/abs/1706.00001}
  {arXiv:1706.00001 [hep-ph]} \BibitemShut {NoStop}%
\bibitem [{\citenamefont {{East}}\ and\ \citenamefont
  {{Lehner}}(2019)}]{EasL19}%
  \BibitemOpen
  \bibfield  {author} {\bibinfo {author} {\bibfnamefont {W.~E.}\ \bibnamefont
  {{East}}}\ and\ \bibinfo {author} {\bibfnamefont {L.}~\bibnamefont
  {{Lehner}}},\ }\bibfield  {title} {\bibinfo {title} {{Fate of a neutron star
  with an endoparasitic black hole and implications for dark matter}},\ }\href
  {https://doi.org/10.1103/PhysRevD.100.124026} {\bibfield  {journal} {\bibinfo
   {journal} {\prd}\ }\textbf {\bibinfo {volume} {100}},\ \bibinfo {eid}
  {124026} (\bibinfo {year} {2019})},\ \Eprint
  {https://arxiv.org/abs/1909.07968} {arXiv:1909.07968 [gr-qc]} \BibitemShut
  {NoStop}%
\bibitem [{\citenamefont {{G{\'e}nolini}}\ \emph {et~al.}(2020)\citenamefont
  {{G{\'e}nolini}}, \citenamefont {{Serpico}},\ and\ \citenamefont
  {{Tinyakov}}}]{GenST20}%
  \BibitemOpen
  \bibfield  {author} {\bibinfo {author} {\bibfnamefont {Y.}~\bibnamefont
  {{G{\'e}nolini}}}, \bibinfo {author} {\bibfnamefont {P.~D.}\ \bibnamefont
  {{Serpico}}},\ and\ \bibinfo {author} {\bibfnamefont {P.}~\bibnamefont
  {{Tinyakov}}},\ }\bibfield  {title} {\bibinfo {title} {{Revisiting primordial
  black hole capture into neutron stars}},\ }\href
  {https://doi.org/10.1103/PhysRevD.102.083004} {\bibfield  {journal} {\bibinfo
   {journal} {\prd}\ }\textbf {\bibinfo {volume} {102}},\ \bibinfo {eid}
  {083004} (\bibinfo {year} {2020})},\ \Eprint
  {https://arxiv.org/abs/2006.16975} {arXiv:2006.16975 [astro-ph.HE]}
  \BibitemShut {NoStop}%
\bibitem [{\citenamefont {{Hawking}}(1971)}]{Haw71}%
  \BibitemOpen
  \bibfield  {author} {\bibinfo {author} {\bibfnamefont {S.}~\bibnamefont
  {{Hawking}}},\ }\bibfield  {title} {\bibinfo {title} {{Gravitationally
  collapsed objects of very low mass}},\ }\href
  {https://doi.org/10.1093/mnras/152.1.75} {\bibfield  {journal} {\bibinfo
  {journal} {\mnras}\ }\textbf {\bibinfo {volume} {152}},\ \bibinfo {pages}
  {75} (\bibinfo {year} {1971})}\BibitemShut {NoStop}%
\bibitem [{\citenamefont {{Carr}}\ and\ \citenamefont
  {{Hawking}}(1974)}]{CarH74}%
  \BibitemOpen
  \bibfield  {author} {\bibinfo {author} {\bibfnamefont {B.~J.}\ \bibnamefont
  {{Carr}}}\ and\ \bibinfo {author} {\bibfnamefont {S.~W.}\ \bibnamefont
  {{Hawking}}},\ }\bibfield  {title} {\bibinfo {title} {{Black holes in the
  early Universe}},\ }\href {https://doi.org/10.1093/mnras/168.2.399}
  {\bibfield  {journal} {\bibinfo  {journal} {\mnras}\ }\textbf {\bibinfo
  {volume} {168}},\ \bibinfo {pages} {399} (\bibinfo {year}
  {1974})}\BibitemShut {NoStop}%
\bibitem [{\citenamefont {{Markovic}}(1995)}]{Mar95}%
  \BibitemOpen
  \bibfield  {author} {\bibinfo {author} {\bibfnamefont {D.}~\bibnamefont
  {{Markovic}}},\ }\bibfield  {title} {\bibinfo {title} {{Evolution of a
  primordial black hole inside a rotating solar-type star}},\ }\href
  {https://doi.org/10.1093/mnras/277.1.25} {\bibfield  {journal} {\bibinfo
  {journal} {\mnras}\ }\textbf {\bibinfo {volume} {277}},\ \bibinfo {pages}
  {25} (\bibinfo {year} {1995})}\BibitemShut {NoStop}%
\bibitem [{\citenamefont {{Montero-Camacho}}\ \emph {et~al.}(2019)\citenamefont
  {{Montero-Camacho}}, \citenamefont {{Fang}}, \citenamefont {{Vasquez}},
  \citenamefont {{Silva}},\ and\ \citenamefont {{Hirata}}}]{MonFVSH19}%
  \BibitemOpen
  \bibfield  {author} {\bibinfo {author} {\bibfnamefont {P.}~\bibnamefont
  {{Montero-Camacho}}}, \bibinfo {author} {\bibfnamefont {X.}~\bibnamefont
  {{Fang}}}, \bibinfo {author} {\bibfnamefont {G.}~\bibnamefont {{Vasquez}}},
  \bibinfo {author} {\bibfnamefont {M.}~\bibnamefont {{Silva}}},\ and\ \bibinfo
  {author} {\bibfnamefont {C.~M.}\ \bibnamefont {{Hirata}}},\ }\bibfield
  {title} {\bibinfo {title} {{Revisiting constraints on asteroid-mass
  primordial black holes as dark matter candidates}},\ }\href
  {https://doi.org/10.1088/1475-7516/2019/08/031} {\bibfield  {journal}
  {\bibinfo  {journal} {\jcap}\ }\textbf {\bibinfo {volume} {2019}},\ \bibinfo
  {eid} {031} (\bibinfo {year} {2019})},\ \Eprint
  {https://arxiv.org/abs/1906.05950} {arXiv:1906.05950 [astro-ph.CO]}
  \BibitemShut {NoStop}%
\bibitem [{\citenamefont {{K{\"u}hnel}}\ and\ \citenamefont
  {{Freese}}(2017)}]{KueF17}%
  \BibitemOpen
  \bibfield  {author} {\bibinfo {author} {\bibfnamefont {F.}~\bibnamefont
  {{K{\"u}hnel}}}\ and\ \bibinfo {author} {\bibfnamefont {K.}~\bibnamefont
  {{Freese}}},\ }\bibfield  {title} {\bibinfo {title} {{Constraints on
  primordial black holes with extended mass functions}},\ }\href
  {https://doi.org/10.1103/PhysRevD.95.083508} {\bibfield  {journal} {\bibinfo
  {journal} {\prd}\ }\textbf {\bibinfo {volume} {95}},\ \bibinfo {eid} {083508}
  (\bibinfo {year} {2017})},\ \Eprint {https://arxiv.org/abs/1701.07223}
  {arXiv:1701.07223 [astro-ph.CO]} \BibitemShut {NoStop}%
\bibitem [{\citenamefont {{Carr}}\ and\ \citenamefont
  {{K{\"u}hnel}}(2020)}]{CarK20}%
  \BibitemOpen
  \bibfield  {author} {\bibinfo {author} {\bibfnamefont {B.}~\bibnamefont
  {{Carr}}}\ and\ \bibinfo {author} {\bibfnamefont {F.}~\bibnamefont
  {{K{\"u}hnel}}},\ }\bibfield  {title} {\bibinfo {title} {{Primordial Black
  Holes as Dark Matter: Recent Developments}},\ }\href@noop {} {\bibfield
  {journal} {\bibinfo  {journal} {Annual Review of Nuclear and Particle
  Science}\ }\textbf {\bibinfo {volume} {70}} (\bibinfo {year} {2020})},\
  \Eprint {https://arxiv.org/abs/2006.02838} {arXiv:2006.02838 [astro-ph.CO]}
  \BibitemShut {NoStop}%
\bibitem [{\citenamefont {{Carr}}\ \emph {et~al.}(2020)\citenamefont {{Carr}},
  \citenamefont {{Kohri}}, \citenamefont {{Sendouda}},\ and\ \citenamefont
  {{Yokoyama}}}]{CarKSY20}%
  \BibitemOpen
  \bibfield  {author} {\bibinfo {author} {\bibfnamefont {B.}~\bibnamefont
  {{Carr}}}, \bibinfo {author} {\bibfnamefont {K.}~\bibnamefont {{Kohri}}},
  \bibinfo {author} {\bibfnamefont {Y.}~\bibnamefont {{Sendouda}}},\ and\
  \bibinfo {author} {\bibfnamefont {J.}~\bibnamefont {{Yokoyama}}},\ }\bibfield
   {title} {\bibinfo {title} {{Constraints on Primordial Black Holes}},\
  }\Eprint {https://arxiv.org/abs/arXiv:2002.12778} {arXiv:2002.12778
  [astro-ph.CO]}  (\bibinfo {year} {2020})\BibitemShut {NoStop}%
\bibitem [{\citenamefont {{Sasaki}}\ \emph {et~al.}(2018)\citenamefont
  {{Sasaki}}, \citenamefont {{Suyama}}, \citenamefont {{Tanaka}},\ and\
  \citenamefont {{Yokoyama}}}]{SasSTY18}%
  \BibitemOpen
  \bibfield  {author} {\bibinfo {author} {\bibfnamefont {M.}~\bibnamefont
  {{Sasaki}}}, \bibinfo {author} {\bibfnamefont {T.}~\bibnamefont {{Suyama}}},
  \bibinfo {author} {\bibfnamefont {T.}~\bibnamefont {{Tanaka}}},\ and\
  \bibinfo {author} {\bibfnamefont {S.}~\bibnamefont {{Yokoyama}}},\ }\bibfield
   {title} {\bibinfo {title} {{Primordial black holes{\textemdash}perspectives
  in gravitational wave astronomy}},\ }\href
  {https://doi.org/10.1088/1361-6382/aaa7b4} {\bibfield  {journal} {\bibinfo
  {journal} {Classical and Quantum Gravity}\ }\textbf {\bibinfo {volume}
  {35}},\ \bibinfo {eid} {063001} (\bibinfo {year} {2018})},\ \Eprint
  {https://arxiv.org/abs/1801.05235} {arXiv:1801.05235 [astro-ph.CO]}
  \BibitemShut {NoStop}%
\bibitem [{\citenamefont {{Vaskonen}}\ and\ \citenamefont
  {{Veerm{\"a}e}}(2021)}]{VasV21}%
  \BibitemOpen
  \bibfield  {author} {\bibinfo {author} {\bibfnamefont {V.}~\bibnamefont
  {{Vaskonen}}}\ and\ \bibinfo {author} {\bibfnamefont {H.}~\bibnamefont
  {{Veerm{\"a}e}}},\ }\bibfield  {title} {\bibinfo {title} {{Did NANOGrav see a
  signal from primordial black hole formation?}},\ }\href@noop {} {\bibfield
  {journal} {\bibinfo  {journal} {\prl}\ }\textbf {\bibinfo {volume} {126}},\
  \bibinfo {eid} {051303} (\bibinfo {year} {2021})},\ \Eprint
  {https://arxiv.org/abs/arXiv:2009.07832} {arXiv:2009.07832 [astro-ph.CO]}
  \BibitemShut {NoStop}%
\bibitem [{\citenamefont {{Fuller}}\ \emph {et~al.}(2017)\citenamefont
  {{Fuller}}, \citenamefont {{Kusenko}},\ and\ \citenamefont
  {{Takhistov}}}]{FulKT17}%
  \BibitemOpen
  \bibfield  {author} {\bibinfo {author} {\bibfnamefont {G.~M.}\ \bibnamefont
  {{Fuller}}}, \bibinfo {author} {\bibfnamefont {A.}~\bibnamefont
  {{Kusenko}}},\ and\ \bibinfo {author} {\bibfnamefont {V.}~\bibnamefont
  {{Takhistov}}},\ }\bibfield  {title} {\bibinfo {title} {{Primordial Black
  Holes and r -Process Nucleosynthesis}},\ }\href
  {https://doi.org/10.1103/PhysRevLett.119.061101} {\bibfield  {journal}
  {\bibinfo  {journal} {\prl}\ }\textbf {\bibinfo {volume} {119}},\ \bibinfo
  {eid} {061101} (\bibinfo {year} {2017})},\ \Eprint
  {https://arxiv.org/abs/1704.01129} {arXiv:1704.01129 [astro-ph.HE]}
  \BibitemShut {NoStop}%
\bibitem [{\citenamefont {{Takhistov}}\ \emph {et~al.}(2020)\citenamefont
  {{Takhistov}}, \citenamefont {{Fuller}},\ and\ \citenamefont
  {{Kusenko}}}]{TakFK20}%
  \BibitemOpen
  \bibfield  {author} {\bibinfo {author} {\bibfnamefont {V.}~\bibnamefont
  {{Takhistov}}}, \bibinfo {author} {\bibfnamefont {G.~M.}\ \bibnamefont
  {{Fuller}}},\ and\ \bibinfo {author} {\bibfnamefont {A.}~\bibnamefont
  {{Kusenko}}},\ }\bibfield  {title} {\bibinfo {title} {{A Test for the Origin
  of Solar Mass Black Holes}},\ }\Eprint {https://arxiv.org/abs/2008.12780}
  {arXiv:2008.12780 [astro-ph.HE]}  (\bibinfo {year} {2020})\BibitemShut
  {NoStop}%
\bibitem [{\citenamefont {{Bondi}}(1952)}]{Bon52}%
  \BibitemOpen
  \bibfield  {author} {\bibinfo {author} {\bibfnamefont {H.}~\bibnamefont
  {{Bondi}}},\ }\bibfield  {title} {\bibinfo {title} {{On spherically
  symmetrical accretion}},\ }\href {https://doi.org/10.1093/mnras/112.2.195}
  {\bibfield  {journal} {\bibinfo  {journal} {\mnras}\ }\textbf {\bibinfo
  {volume} {112}},\ \bibinfo {pages} {195} (\bibinfo {year}
  {1952})}\BibitemShut {NoStop}%
\bibitem [{\citenamefont {{Shapiro}}\ and\ \citenamefont
  {{Teukolsky}}(2004)}]{Shapiro}%
  \BibitemOpen
  \bibfield  {author} {\bibinfo {author} {\bibfnamefont {S.~L.}\ \bibnamefont
  {{Shapiro}}}\ and\ \bibinfo {author} {\bibfnamefont {S.~A.}\ \bibnamefont
  {{Teukolsky}}},\ }\href@noop {} {\emph {\bibinfo {title} {{Black Holes, White
  Dwarfs, and Neutron Stars: The Physics of Compact Objects}}}}\ (\bibinfo
  {publisher} {Wiley-VCH, New York},\ \bibinfo {year} {2004})\BibitemShut
  {NoStop}%
\bibitem [{\citenamefont {{Kouvaris}}\ and\ \citenamefont
  {{Tinyakov}}(2014)}]{KouT14}%
  \BibitemOpen
  \bibfield  {author} {\bibinfo {author} {\bibfnamefont {C.}~\bibnamefont
  {{Kouvaris}}}\ and\ \bibinfo {author} {\bibfnamefont {P.}~\bibnamefont
  {{Tinyakov}}},\ }\bibfield  {title} {\bibinfo {title} {{Growth of black holes
  in the interior of rotating neutron stars}},\ }\href
  {https://doi.org/10.1103/PhysRevD.90.043512} {\bibfield  {journal} {\bibinfo
  {journal} {\prd}\ }\textbf {\bibinfo {volume} {90}},\ \bibinfo {eid} {043512}
  (\bibinfo {year} {2014})},\ \Eprint {https://arxiv.org/abs/1312.3764}
  {arXiv:1312.3764 [astro-ph.SR]} \BibitemShut {NoStop}%
\bibitem [{\citenamefont {{Richards}}\ \emph {et~al.}(2021)\citenamefont
  {{Richards}}, \citenamefont {{Baumgarte}},\ and\ \citenamefont
  {{Shapiro}}}]{RicBS21a}%
  \BibitemOpen
  \bibfield  {author} {\bibinfo {author} {\bibfnamefont {C.~B.}\ \bibnamefont
  {{Richards}}}, \bibinfo {author} {\bibfnamefont {T.~W.}\ \bibnamefont
  {{Baumgarte}}},\ and\ \bibinfo {author} {\bibfnamefont {S.~L.}\ \bibnamefont
  {{Shapiro}}},\ }\bibfield  {title} {\bibinfo {title} {{Relativistic Bondi
  accretion for stiff equations of state}},\ }\href@noop {} {\bibfield
  {journal} {\bibinfo  {journal} {\mnras}\ }\textbf {\bibinfo {volume} {502}},\
  \bibinfo {eid} {3003} (\bibinfo {year} {2021})},\ \Eprint
  {https://arxiv.org/abs/2101.08797} {arXiv:2101.08797 [astro-ph.HE]}
  \BibitemShut {NoStop}%
\bibitem [{\citenamefont {{Baumgarte}}\ and\ \citenamefont
  {{Shapiro}}(2021)}]{BauS21}%
  \BibitemOpen
  \bibfield  {author} {\bibinfo {author} {\bibfnamefont {T.~W.}\ \bibnamefont
  {{Baumgarte}}}\ and\ \bibinfo {author} {\bibfnamefont {S.~L.}\ \bibnamefont
  {{Shapiro}}},\ }\bibfield  {title} {\bibinfo {title} {{Neutron stars
  harboring a primordial black hole: Maximum survival time}},\ }\Eprint
  {https://arxiv.org/abs/2101.12220} {arXiv:2101.12220 [astro-ph.HE]}
  (\bibinfo {year} {2021})\BibitemShut {NoStop}%
\bibitem [{\citenamefont {{Baumgarte}}\ \emph {et~al.}(2013)\citenamefont
  {{Baumgarte}}, \citenamefont {{Montero}}, \citenamefont
  {{Cordero-Carri{\'o}n}},\ and\ \citenamefont {{M{\"u}ller}}}]{BauMCM13}%
  \BibitemOpen
  \bibfield  {author} {\bibinfo {author} {\bibfnamefont {T.~W.}\ \bibnamefont
  {{Baumgarte}}}, \bibinfo {author} {\bibfnamefont {P.~J.}\ \bibnamefont
  {{Montero}}}, \bibinfo {author} {\bibfnamefont {I.}~\bibnamefont
  {{Cordero-Carri{\'o}n}}},\ and\ \bibinfo {author} {\bibfnamefont
  {E.}~\bibnamefont {{M{\"u}ller}}},\ }\bibfield  {title} {\bibinfo {title}
  {{Numerical relativity in spherical polar coordinates: Evolution calculations
  with the BSSN formulation}},\ }\href
  {https://doi.org/10.1103/PhysRevD.87.044026} {\bibfield  {journal} {\bibinfo
  {journal} {\prd}\ }\textbf {\bibinfo {volume} {87}},\ \bibinfo {eid} {044026}
  (\bibinfo {year} {2013})},\ \Eprint {https://arxiv.org/abs/1211.6632}
  {arXiv:1211.6632 [gr-qc]} \BibitemShut {NoStop}%
\bibitem [{\citenamefont {{Baumgarte}}\ \emph {et~al.}(2015)\citenamefont
  {{Baumgarte}}, \citenamefont {{Montero}},\ and\ \citenamefont
  {{M{\"u}ller}}}]{BauMM15}%
  \BibitemOpen
  \bibfield  {author} {\bibinfo {author} {\bibfnamefont {T.~W.}\ \bibnamefont
  {{Baumgarte}}}, \bibinfo {author} {\bibfnamefont {P.~J.}\ \bibnamefont
  {{Montero}}},\ and\ \bibinfo {author} {\bibfnamefont {E.}~\bibnamefont
  {{M{\"u}ller}}},\ }\bibfield  {title} {\bibinfo {title} {{Numerical
  relativity in spherical polar coordinates: Off-center simulations}},\ }\href
  {https://doi.org/10.1103/PhysRevD.91.064035} {\bibfield  {journal} {\bibinfo
  {journal} {\prd}\ }\textbf {\bibinfo {volume} {91}},\ \bibinfo {eid} {064035}
  (\bibinfo {year} {2015})},\ \Eprint {https://arxiv.org/abs/1501.05259}
  {arXiv:1501.05259 [gr-qc]} \BibitemShut {NoStop}%
\bibitem [{\citenamefont {{Michel}}(1972)}]{Mic72}%
  \BibitemOpen
  \bibfield  {author} {\bibinfo {author} {\bibfnamefont {F.~C.}\ \bibnamefont
  {{Michel}}},\ }\bibfield  {title} {\bibinfo {title} {{Accretion of Matter by
  Condensed Objects}},\ }\href {https://doi.org/10.1007/BF00649949} {\bibfield
  {journal} {\bibinfo  {journal} {\apss}\ }\textbf {\bibinfo {volume} {15}},\
  \bibinfo {pages} {153} (\bibinfo {year} {1972})}\BibitemShut {NoStop}%
\bibitem [{\citenamefont {{Tolman}}(1939)}]{Tol39}%
  \BibitemOpen
  \bibfield  {author} {\bibinfo {author} {\bibfnamefont {R.~C.}\ \bibnamefont
  {{Tolman}}},\ }\bibfield  {title} {\bibinfo {title} {{Static Solutions of
  Einstein's Field Equations for Spheres of Fluid}},\ }\href
  {https://doi.org/10.1103/PhysRev.55.364} {\bibfield  {journal} {\bibinfo
  {journal} {Physical Review}\ }\textbf {\bibinfo {volume} {55}},\ \bibinfo
  {pages} {364} (\bibinfo {year} {1939})}\BibitemShut {NoStop}%
\bibitem [{\citenamefont {{Oppenheimer}}\ and\ \citenamefont
  {{Volkoff}}(1939)}]{OppV39}%
  \BibitemOpen
  \bibfield  {author} {\bibinfo {author} {\bibfnamefont {J.~R.}\ \bibnamefont
  {{Oppenheimer}}}\ and\ \bibinfo {author} {\bibfnamefont {G.~M.}\ \bibnamefont
  {{Volkoff}}},\ }\bibfield  {title} {\bibinfo {title} {{On Massive Neutron
  Cores}},\ }\href {https://doi.org/10.1103/PhysRev.55.374} {\bibfield
  {journal} {\bibinfo  {journal} {Physical Review}\ }\textbf {\bibinfo {volume}
  {55}},\ \bibinfo {pages} {374} (\bibinfo {year} {1939})}\BibitemShut
  {NoStop}%
\bibitem [{\citenamefont {{Baumgarte}}\ and\ \citenamefont
  {{Shapiro}}(2010)}]{BauS10}%
  \BibitemOpen
  \bibfield  {author} {\bibinfo {author} {\bibfnamefont {T.~W.}\ \bibnamefont
  {{Baumgarte}}}\ and\ \bibinfo {author} {\bibfnamefont {S.~L.}\ \bibnamefont
  {{Shapiro}}},\ }\href@noop {} {\emph {\bibinfo {title} {{Numerical
  Relativity: Solving Einstein's Equations on the Computer}}}}\ (\bibinfo
  {publisher} {Cambridge University Press},\ \bibinfo {year}
  {2010})\BibitemShut {NoStop}%
\bibitem [{\citenamefont {{Brandt}}\ and\ \citenamefont
  {{Br{\"u}gmann}}(1997)}]{BraB97}%
  \BibitemOpen
  \bibfield  {author} {\bibinfo {author} {\bibfnamefont {S.}~\bibnamefont
  {{Brandt}}}\ and\ \bibinfo {author} {\bibfnamefont {B.}~\bibnamefont
  {{Br{\"u}gmann}}},\ }\bibfield  {title} {\bibinfo {title} {{A Simple
  Construction of Initial Data for Multiple Black Holes}},\ }\href
  {https://doi.org/10.1103/PhysRevLett.78.3606} {\bibfield  {journal} {\bibinfo
   {journal} {\prl}\ }\textbf {\bibinfo {volume} {78}},\ \bibinfo {pages}
  {3606} (\bibinfo {year} {1997})},\ \Eprint
  {https://arxiv.org/abs/gr-qc/9703066} {arXiv:gr-qc/9703066 [gr-qc]}
  \BibitemShut {NoStop}%
\bibitem [{\citenamefont {{Nakamura}}\ \emph {et~al.}(1987)\citenamefont
  {{Nakamura}}, \citenamefont {{Oohara}},\ and\ \citenamefont
  {{Kojima}}}]{NakOK87}%
  \BibitemOpen
  \bibfield  {author} {\bibinfo {author} {\bibfnamefont {T.}~\bibnamefont
  {{Nakamura}}}, \bibinfo {author} {\bibfnamefont {K.}~\bibnamefont
  {{Oohara}}},\ and\ \bibinfo {author} {\bibfnamefont {Y.}~\bibnamefont
  {{Kojima}}},\ }\bibfield  {title} {\bibinfo {title} {{General Relativistic
  Collapse to Black Holes and Gravitational Waves from Black Holes}},\ }\href
  {https://doi.org/10.1143/PTPS.90.1} {\bibfield  {journal} {\bibinfo
  {journal} {Progress of Theoretical Physics Supplement}\ }\textbf {\bibinfo
  {volume} {90}},\ \bibinfo {pages} {1} (\bibinfo {year} {1987})}\BibitemShut
  {NoStop}%
\bibitem [{\citenamefont {{Shibata}}\ and\ \citenamefont
  {{Nakamura}}(1995)}]{ShiN95}%
  \BibitemOpen
  \bibfield  {author} {\bibinfo {author} {\bibfnamefont {M.}~\bibnamefont
  {{Shibata}}}\ and\ \bibinfo {author} {\bibfnamefont {T.}~\bibnamefont
  {{Nakamura}}},\ }\bibfield  {title} {\bibinfo {title} {{Evolution of
  three-dimensional gravitational waves: Harmonic slicing case}},\ }\href
  {https://doi.org/10.1103/PhysRevD.52.5428} {\bibfield  {journal} {\bibinfo
  {journal} {\prd}\ }\textbf {\bibinfo {volume} {52}},\ \bibinfo {pages} {5428}
  (\bibinfo {year} {1995})}\BibitemShut {NoStop}%
\bibitem [{\citenamefont {{Baumgarte}}\ and\ \citenamefont
  {{Shapiro}}(1999)}]{BauS99}%
  \BibitemOpen
  \bibfield  {author} {\bibinfo {author} {\bibfnamefont {T.~W.}\ \bibnamefont
  {{Baumgarte}}}\ and\ \bibinfo {author} {\bibfnamefont {S.~L.}\ \bibnamefont
  {{Shapiro}}},\ }\bibfield  {title} {\bibinfo {title} {{Numerical integration
  of Einstein's field equations}},\ }\href
  {https://doi.org/10.1103/PhysRevD.59.024007} {\bibfield  {journal} {\bibinfo
  {journal} {\prd}\ }\textbf {\bibinfo {volume} {59}},\ \bibinfo {eid} {024007}
  (\bibinfo {year} {1999})},\ \Eprint {https://arxiv.org/abs/gr-qc/9810065}
  {arXiv:gr-qc/9810065 [gr-qc]} \BibitemShut {NoStop}%
\bibitem [{\citenamefont {{Bonazzola}}\ \emph {et~al.}(2004)\citenamefont
  {{Bonazzola}}, \citenamefont {{Gourgoulhon}}, \citenamefont
  {{Grandcl{\'e}ment}},\ and\ \citenamefont {{Novak}}}]{BonGGN04}%
  \BibitemOpen
  \bibfield  {author} {\bibinfo {author} {\bibfnamefont {S.}~\bibnamefont
  {{Bonazzola}}}, \bibinfo {author} {\bibfnamefont {E.}~\bibnamefont
  {{Gourgoulhon}}}, \bibinfo {author} {\bibfnamefont {P.}~\bibnamefont
  {{Grandcl{\'e}ment}}},\ and\ \bibinfo {author} {\bibfnamefont
  {J.}~\bibnamefont {{Novak}}},\ }\bibfield  {title} {\bibinfo {title}
  {{Constrained scheme for the Einstein equations based on the Dirac gauge and
  spherical coordinates}},\ }\href {https://doi.org/10.1103/PhysRevD.70.104007}
  {\bibfield  {journal} {\bibinfo  {journal} {\prd}\ }\textbf {\bibinfo
  {volume} {70}},\ \bibinfo {eid} {104007} (\bibinfo {year} {2004})},\ \Eprint
  {https://arxiv.org/abs/gr-qc/0307082} {arXiv:gr-qc/0307082 [gr-qc]}
  \BibitemShut {NoStop}%
\bibitem [{\citenamefont {{Shibata}}\ \emph {et~al.}(2004)\citenamefont
  {{Shibata}}, \citenamefont {{Ury{\={u}}}},\ and\ \citenamefont
  {{Friedman}}}]{ShiUF04}%
  \BibitemOpen
  \bibfield  {author} {\bibinfo {author} {\bibfnamefont {M.}~\bibnamefont
  {{Shibata}}}, \bibinfo {author} {\bibfnamefont {K.}~\bibnamefont
  {{Ury{\={u}}}}},\ and\ \bibinfo {author} {\bibfnamefont {J.~L.}\ \bibnamefont
  {{Friedman}}},\ }\bibfield  {title} {\bibinfo {title} {{Deriving formulations
  for numerical computation of binary neutron stars in quasicircular orbits}},\
  }\href {https://doi.org/10.1103/PhysRevD.70.044044} {\bibfield  {journal}
  {\bibinfo  {journal} {\prd}\ }\textbf {\bibinfo {volume} {70}},\ \bibinfo
  {eid} {044044} (\bibinfo {year} {2004})},\ \Eprint
  {https://arxiv.org/abs/gr-qc/0407036} {arXiv:gr-qc/0407036 [gr-qc]}
  \BibitemShut {NoStop}%
\bibitem [{\citenamefont {{Brown}}(2009)}]{Bro09}%
  \BibitemOpen
  \bibfield  {author} {\bibinfo {author} {\bibfnamefont {J.~D.}\ \bibnamefont
  {{Brown}}},\ }\bibfield  {title} {\bibinfo {title} {{Covariant formulations
  of Baumgarte, Shapiro, Shibata, and Nakamura and the standard gauge}},\
  }\href {https://doi.org/10.1103/PhysRevD.79.104029} {\bibfield  {journal}
  {\bibinfo  {journal} {\prd}\ }\textbf {\bibinfo {volume} {79}},\ \bibinfo
  {eid} {104029} (\bibinfo {year} {2009})},\ \Eprint
  {https://arxiv.org/abs/0902.3652} {arXiv:0902.3652 [gr-qc]} \BibitemShut
  {NoStop}%
\bibitem [{\citenamefont {{Gourgoulhon}}(2012)}]{Gou12}%
  \BibitemOpen
  \bibfield  {author} {\bibinfo {author} {\bibfnamefont {E.}~\bibnamefont
  {{Gourgoulhon}}},\ }\href {https://doi.org/10.1007/978-3-642-24525-1} {\emph
  {\bibinfo {title} {{3+1 Formalism in General Relativity}}}}\ (\bibinfo
  {publisher} {{Springer, Berlin}},\ \bibinfo {year} {2012})\BibitemShut
  {NoStop}%
\bibitem [{\citenamefont {{Miller}}\ and\ \citenamefont
  {{Baumgarte}}(2017)}]{MilB17}%
  \BibitemOpen
  \bibfield  {author} {\bibinfo {author} {\bibfnamefont {A.~J.}\ \bibnamefont
  {{Miller}}}\ and\ \bibinfo {author} {\bibfnamefont {T.~W.}\ \bibnamefont
  {{Baumgarte}}},\ }\bibfield  {title} {\bibinfo {title} {{Bondi accretion in
  trumpet geometries}},\ }\href {https://doi.org/10.1088/1361-6382/aa51fe}
  {\bibfield  {journal} {\bibinfo  {journal} {Classical and Quantum Gravity}\
  }\textbf {\bibinfo {volume} {34}},\ \bibinfo {eid} {035007} (\bibinfo {year}
  {2017})},\ \Eprint {https://arxiv.org/abs/1607.03047} {arXiv:1607.03047
  [gr-qc]} \BibitemShut {NoStop}%
\bibitem [{\citenamefont {{Ruchlin}}\ \emph {et~al.}(2018)\citenamefont
  {{Ruchlin}}, \citenamefont {{Etienne}},\ and\ \citenamefont
  {{Baumgarte}}}]{RucEB18}%
  \BibitemOpen
  \bibfield  {author} {\bibinfo {author} {\bibfnamefont {I.}~\bibnamefont
  {{Ruchlin}}}, \bibinfo {author} {\bibfnamefont {Z.~B.}\ \bibnamefont
  {{Etienne}}},\ and\ \bibinfo {author} {\bibfnamefont {T.~W.}\ \bibnamefont
  {{Baumgarte}}},\ }\bibfield  {title} {\bibinfo {title} {{SENR /NRPy + :
  Numerical relativity in singular curvilinear coordinate systems}},\ }\href
  {https://doi.org/10.1103/PhysRevD.97.064036} {\bibfield  {journal} {\bibinfo
  {journal} {\prd}\ }\textbf {\bibinfo {volume} {97}},\ \bibinfo {eid} {064036}
  (\bibinfo {year} {2018})},\ \Eprint {https://arxiv.org/abs/1712.07658}
  {arXiv:1712.07658 [gr-qc]} \BibitemShut {NoStop}%
\bibitem [{\citenamefont {{Mewes}}\ \emph {et~al.}(2018)\citenamefont
  {{Mewes}}, \citenamefont {{Zlochower}}, \citenamefont {{Campanelli}},
  \citenamefont {{Ruchlin}}, \citenamefont {{Etienne}},\ and\ \citenamefont
  {{Baumgarte}}}]{MewZCREB18}%
  \BibitemOpen
  \bibfield  {author} {\bibinfo {author} {\bibfnamefont {V.}~\bibnamefont
  {{Mewes}}}, \bibinfo {author} {\bibfnamefont {Y.}~\bibnamefont
  {{Zlochower}}}, \bibinfo {author} {\bibfnamefont {M.}~\bibnamefont
  {{Campanelli}}}, \bibinfo {author} {\bibfnamefont {I.}~\bibnamefont
  {{Ruchlin}}}, \bibinfo {author} {\bibfnamefont {Z.~B.}\ \bibnamefont
  {{Etienne}}},\ and\ \bibinfo {author} {\bibfnamefont {T.~W.}\ \bibnamefont
  {{Baumgarte}}},\ }\bibfield  {title} {\bibinfo {title} {{Numerical relativity
  in spherical coordinates with the Einstein Toolkit}},\ }\href
  {https://doi.org/10.1103/PhysRevD.97.084059} {\bibfield  {journal} {\bibinfo
  {journal} {\prd}\ }\textbf {\bibinfo {volume} {97}},\ \bibinfo {eid} {084059}
  (\bibinfo {year} {2018})},\ \Eprint {https://arxiv.org/abs/1802.09625}
  {arXiv:1802.09625 [gr-qc]} \BibitemShut {NoStop}%
\bibitem [{\citenamefont {{Mewes}}\ \emph {et~al.}(2020)\citenamefont
  {{Mewes}}, \citenamefont {{Zlochower}}, \citenamefont {{Campanelli}},
  \citenamefont {{Baumgarte}}, \citenamefont {{Etienne}}, \citenamefont
  {{Armengol}},\ and\ \citenamefont {{Cipolletta}}}]{MewZCBEAC20}%
  \BibitemOpen
  \bibfield  {author} {\bibinfo {author} {\bibfnamefont {V.}~\bibnamefont
  {{Mewes}}}, \bibinfo {author} {\bibfnamefont {Y.}~\bibnamefont
  {{Zlochower}}}, \bibinfo {author} {\bibfnamefont {M.}~\bibnamefont
  {{Campanelli}}}, \bibinfo {author} {\bibfnamefont {T.~W.}\ \bibnamefont
  {{Baumgarte}}}, \bibinfo {author} {\bibfnamefont {Z.~B.}\ \bibnamefont
  {{Etienne}}}, \bibinfo {author} {\bibfnamefont {F.~G.~L.}\ \bibnamefont
  {{Armengol}}},\ and\ \bibinfo {author} {\bibfnamefont {F.}~\bibnamefont
  {{Cipolletta}}},\ }\bibfield  {title} {\bibinfo {title} {{Numerical
  relativity in spherical coordinates: A new dynamical spacetime and general
  relativistic MHD evolution framework for the Einstein Toolkit}},\ }\href
  {https://doi.org/10.1103/PhysRevD.101.104007} {\bibfield  {journal} {\bibinfo
   {journal} {\prd}\ }\textbf {\bibinfo {volume} {101}},\ \bibinfo {eid}
  {104007} (\bibinfo {year} {2020})},\ \Eprint
  {https://arxiv.org/abs/2002.06225} {arXiv:2002.06225 [gr-qc]} \BibitemShut
  {NoStop}%
\bibitem [{\citenamefont {{Bona}}\ \emph {et~al.}(1995)\citenamefont {{Bona}},
  \citenamefont {{Mass{\'o}}}, \citenamefont {{Seidel}},\ and\ \citenamefont
  {{Stela}}}]{BonMSS95}%
  \BibitemOpen
  \bibfield  {author} {\bibinfo {author} {\bibfnamefont {C.}~\bibnamefont
  {{Bona}}}, \bibinfo {author} {\bibfnamefont {J.}~\bibnamefont {{Mass{\'o}}}},
  \bibinfo {author} {\bibfnamefont {E.}~\bibnamefont {{Seidel}}},\ and\
  \bibinfo {author} {\bibfnamefont {J.}~\bibnamefont {{Stela}}},\ }\bibfield
  {title} {\bibinfo {title} {{New Formalism for Numerical Relativity}},\
  }\href@noop {} {\bibfield  {journal} {\bibinfo  {journal} {\prl}\ }\textbf
  {\bibinfo {volume} {75}},\ \bibinfo {pages} {600} (\bibinfo {year}
  {1995})}\BibitemShut {NoStop}%
\bibitem [{\citenamefont {{Alcubierre}}\ \emph {et~al.}(2003)\citenamefont
  {{Alcubierre}}, \citenamefont {{Br{\"u}gmann}}, \citenamefont {{Diener}},
  \citenamefont {{Koppitz}}, \citenamefont {{Pollney}}, \citenamefont
  {{Seidel}},\ and\ \citenamefont {{Takahashi}}}]{Alcetal03}%
  \BibitemOpen
  \bibfield  {author} {\bibinfo {author} {\bibfnamefont {M.}~\bibnamefont
  {{Alcubierre}}}, \bibinfo {author} {\bibfnamefont {B.}~\bibnamefont
  {{Br{\"u}gmann}}}, \bibinfo {author} {\bibfnamefont {P.}~\bibnamefont
  {{Diener}}}, \bibinfo {author} {\bibfnamefont {M.}~\bibnamefont {{Koppitz}}},
  \bibinfo {author} {\bibfnamefont {D.}~\bibnamefont {{Pollney}}}, \bibinfo
  {author} {\bibfnamefont {E.}~\bibnamefont {{Seidel}}},\ and\ \bibinfo
  {author} {\bibfnamefont {R.}~\bibnamefont {{Takahashi}}},\ }\bibfield
  {title} {\bibinfo {title} {{Gauge conditions for long-term numerical black
  hole evolutions without excision}},\ }\href
  {https://doi.org/10.1103/PhysRevD.67.084023} {\bibfield  {journal} {\bibinfo
  {journal} {\prd}\ }\textbf {\bibinfo {volume} {67}},\ \bibinfo {eid} {084023}
  (\bibinfo {year} {2003})},\ \Eprint {https://arxiv.org/abs/gr-qc/0206072}
  {arXiv:gr-qc/0206072 [gr-qc]} \BibitemShut {NoStop}%
\bibitem [{\citenamefont {{Thierfelder}}\ \emph {et~al.}(2011)\citenamefont
  {{Thierfelder}}, \citenamefont {{Bernuzzi}},\ and\ \citenamefont
  {{Br{\"u}gmann}}}]{ThiBB11}%
  \BibitemOpen
  \bibfield  {author} {\bibinfo {author} {\bibfnamefont {M.}~\bibnamefont
  {{Thierfelder}}}, \bibinfo {author} {\bibfnamefont {S.}~\bibnamefont
  {{Bernuzzi}}},\ and\ \bibinfo {author} {\bibfnamefont {B.}~\bibnamefont
  {{Br{\"u}gmann}}},\ }\bibfield  {title} {\bibinfo {title} {{Numerical
  relativity simulations of binary neutron stars}},\ }\href
  {https://doi.org/10.1103/PhysRevD.84.044012} {\bibfield  {journal} {\bibinfo
  {journal} {\prd}\ }\textbf {\bibinfo {volume} {84}},\ \bibinfo {eid} {044012}
  (\bibinfo {year} {2011})},\ \Eprint {https://arxiv.org/abs/1104.4751}
  {arXiv:1104.4751 [gr-qc]} \BibitemShut {NoStop}%
\bibitem [{\citenamefont {{Montero}}\ \emph {et~al.}(2014)\citenamefont
  {{Montero}}, \citenamefont {{Baumgarte}},\ and\ \citenamefont
  {{M{\"u}ller}}}]{MonBM14}%
  \BibitemOpen
  \bibfield  {author} {\bibinfo {author} {\bibfnamefont {P.~J.}\ \bibnamefont
  {{Montero}}}, \bibinfo {author} {\bibfnamefont {T.~W.}\ \bibnamefont
  {{Baumgarte}}},\ and\ \bibinfo {author} {\bibfnamefont {E.}~\bibnamefont
  {{M{\"u}ller}}},\ }\bibfield  {title} {\bibinfo {title} {{General
  relativistic hydrodynamics in curvilinear coordinates}},\ }\href
  {https://doi.org/10.1103/PhysRevD.89.084043} {\bibfield  {journal} {\bibinfo
  {journal} {\prd}\ }\textbf {\bibinfo {volume} {89}},\ \bibinfo {eid} {084043}
  (\bibinfo {year} {2014})},\ \Eprint {https://arxiv.org/abs/1309.7808}
  {arXiv:1309.7808 [gr-qc]} \BibitemShut {NoStop}%
\bibitem [{\citenamefont {{Harten}}\ \emph {et~al.}(1983)\citenamefont
  {{Harten}}, \citenamefont {{Lax}},\ and\ \citenamefont {{Leer}}}]{HarLL83}%
  \BibitemOpen
  \bibfield  {author} {\bibinfo {author} {\bibfnamefont {A.}~\bibnamefont
  {{Harten}}}, \bibinfo {author} {\bibfnamefont {P.~D.}\ \bibnamefont
  {{Lax}}},\ and\ \bibinfo {author} {\bibfnamefont {v.~B.}\ \bibnamefont
  {{Leer}}},\ }\bibfield  {title} {\bibinfo {title} {{On upstream differencing
  and Godunov type methods for hyperbolic conservation laws}},\ }\href@noop {}
  {\bibfield  {journal} {\bibinfo  {journal} {SIAM Rev.}\ }\textbf {\bibinfo
  {volume} {25}},\ \bibinfo {pages} {35} (\bibinfo {year} {1983})}\BibitemShut
  {NoStop}%
\bibitem [{\citenamefont {{Einfeldt}}(1988)}]{Ein88}%
  \BibitemOpen
  \bibfield  {author} {\bibinfo {author} {\bibfnamefont {B.}~\bibnamefont
  {{Einfeldt}}},\ }\bibfield  {title} {\bibinfo {title} {On godunov methods for
  gas dynamics},\ }\href@noop {} {\bibfield  {journal} {\bibinfo  {journal}
  {SIAM J. Numer. Anal.}\ }\textbf {\bibinfo {volume} {25}},\ \bibinfo {pages}
  {294} (\bibinfo {year} {1988})}\BibitemShut {NoStop}%
\bibitem [{\citenamefont {{van Leer}}(1977)}]{Van77}%
  \BibitemOpen
  \bibfield  {author} {\bibinfo {author} {\bibfnamefont {B.}~\bibnamefont {{van
  Leer}}},\ }\bibfield  {title} {\bibinfo {title} {{Towards the ultimate
  conservative difference scheme: IV. A new approach to numerical
  convection}},\ }\href@noop {} {\bibfield  {journal} {\bibinfo  {journal}
  {Journal of Computational Physics}\ }\textbf {\bibinfo {volume} {23}},\
  \bibinfo {pages} {276} (\bibinfo {year} {1977})}\BibitemShut {NoStop}%
\bibitem [{\citenamefont {{Farris}}\ \emph {et~al.}(2010)\citenamefont
  {{Farris}}, \citenamefont {{Liu}},\ and\ \citenamefont
  {{Shapiro}}}]{FarLS10}%
  \BibitemOpen
  \bibfield  {author} {\bibinfo {author} {\bibfnamefont {B.~D.}\ \bibnamefont
  {{Farris}}}, \bibinfo {author} {\bibfnamefont {Y.~T.}\ \bibnamefont
  {{Liu}}},\ and\ \bibinfo {author} {\bibfnamefont {S.~L.}\ \bibnamefont
  {{Shapiro}}},\ }\bibfield  {title} {\bibinfo {title} {{Binary black hole
  mergers in gaseous environments: ``Binary Bondi`` and ``binary
  Bondi-Hoyle-Lyttleton'' accretion}},\ }\href
  {https://doi.org/10.1103/PhysRevD.81.084008} {\bibfield  {journal} {\bibinfo
  {journal} {\prd}\ }\textbf {\bibinfo {volume} {81}},\ \bibinfo {eid} {084008}
  (\bibinfo {year} {2010})},\ \Eprint {https://arxiv.org/abs/0912.2096}
  {arXiv:0912.2096 [astro-ph.HE]} \BibitemShut {NoStop}%
\bibitem [{\citenamefont {{Carroll}}(2004)}]{Car04}%
  \BibitemOpen
  \bibfield  {author} {\bibinfo {author} {\bibfnamefont {S.~M.}\ \bibnamefont
  {{Carroll}}},\ }\href@noop {} {\emph {\bibinfo {title} {{Spacetime and
  Geometry. An Introduction to General Relativity}}}}\ (\bibinfo  {publisher}
  {Addison Wesley, San Francisco},\ \bibinfo {year} {2004})\BibitemShut
  {NoStop}%
\end{thebibliography}

%

\end{document}